\pdfoutput=1

\documentclass{jfm}

\usepackage{graphicx}
\usepackage{amsmath}
\usepackage{amssymb}
\usepackage{bm}
\usepackage{pifont}
\usepackage{graphicx}
\usepackage{subfig}
\usepackage{hyperref}
\usepackage[usenames,dvipsnames,svgnames,table]{xcolor}
\usepackage{natbib}

\ifCUPmtlplainloaded \else
  \checkfont{eurm10}
  \iffontfound
    \IfFileExists{upmath.sty}
      {\typeout{^^JFound AMS Euler Roman fonts on the system,
                   using the 'upmath' package.^^J}%
       \usepackage{upmath}}
      {\typeout{^^JFound AMS Euler Roman fonts on the system, but you
                   dont seem to have the}%
       \typeout{'upmath' package installed. JFM.cls can take advantage
                 of these fonts,^^Jif you use 'upmath' package.^^J}%
      }
  \else
  \fi
\fi

\ifCUPmtlplainloaded \else
  \checkfont{msam10}
  \iffontfound
    \IfFileExists{amssymb.sty}
      {\typeout{^^JFound AMS Symbol fonts on the system, using the
                'amssymb' package.^^J}%
       \usepackage{amssymb}%
       \let\le=\leqslant  
       \let\ge=\geqslant  
      }{}
  \fi
\fi

\ifCUPmtlplainloaded \else
  \IfFileExists{amsbsy.sty}
    {\typeout{^^JFound the 'amsbsy' package on the system, using it.^^J}%
     \usepackage{amsbsy}}
    {}
\fi

\newsavebox{\astrutbox}
\sbox{\astrutbox}{\rule[-5pt]{0pt}{20pt}}

\newcommand\etal{\mbox{\textit{et al.}}}

\DeclareMathOperator{\sign}{sign}

\title[Azimuthal velocities and axial transport in Rayleigh-stable Taylor-Couette]{Azimuthal velocity profiles in Rayleigh-stable Taylor-Couette flow and implied axial angular momentum transport}

\author[F. Nordsiek \etal]%
{Freja Nordsiek$^1$,\ns
Sander G. Huisman$^2$,\ns
Roeland C. A. van der Veen$^2$,\ns
Chao Sun$^2$ \thanks{Email address for correspondence: \href{mailto:c.sun@utwente.nl}{c.sun@utwente.nl}},\ns
Detlef Lohse$^2$\thanks{Email address for correspondence: \href{mailto:d.lohse@utwente.nl}{d.lohse@utwente.nl}},\ns
and Daniel P. Lathrop$^1$\thanks{Email address for correspondence: \href{mailto:lathrop@umd.edu}{lathrop@umd.edu}}}

\affiliation{$^1$Department of Physics and Institute for Research in Electronics and Applied Physics, University of Maryland, College Park, MD 20742, USA\\[\affilskip]
$^2$ Physics of Fluids Group, Faculty of Science and Technology, MESA+ Institute, and Burgers Center for Fluid Dynamics, University of Twente, 7500AE Enschede, The Netherlands}

\pubyear{2010}
\volume{650}
\pagerange{119--126}
\date{?; revised ?; accepted ?. - To be entered by editorial office}
\begin{document}

\maketitle

\begin{abstract}
We present azimuthal velocity profiles measured in a Taylor-Couette apparatus, which has been used as a model of stellar and planetary accretion disks.
The apparatus has a cylinder radius ratio of $\eta = 0.716$, an aspect-ratio of $\Gamma = 11.74$, and the plates closing the cylinders in the axial direction are attached to the outer cylinder.
We investigate angular momentum transport and Ekman pumping in the Rayleigh-stable regime.
The regime is linearly stable and is characterized by radially increasing specific angular momentum.
We present several Rayleigh-stable profiles for shear Reynolds numbers $Re_S \sim O\left(10^5\right) \,$, both for $\Omega_i > \Omega_o > 0$ (quasi-Keplerian regime) and $\Omega_o > \Omega_i > 0$ (sub-rotating regime) where $\Omega_{i,o}$ is the inner/outer cylinder rotation rate.
None of the velocity profiles matches the non-vortical laminar Taylor-Couette profile.
The deviation from that profile increased as solid-body rotation is approached at fixed $Re_S$.
Flow super-rotation, an angular velocity greater than that of both cylinders, is observed in the sub-rotating regime.
The velocity profiles give lower bounds for the torques required to rotate the inner cylinder that were larger than the torques for the case of laminar Taylor-Couette flow.
The quasi-Keplerian profiles are composed of a well mixed inner region, having approximately constant angular momentum, connected to an outer region in solid-body rotation with the outer cylinder and attached axial boundaries.
These regions suggest that the angular momentum is transported axially to the axial boundaries.
Therefore, Taylor-Couette flow with closing plates attached to the outer cylinder is an imperfect model for accretion disk flows, especially with regard to their stability.
\end{abstract}


\section{Introduction}

Rotating shear flows are common in nature.
Geophysical and astrophysical examples include the interiors of planets and stars, planetary atmospheres, and stellar and planetary accretion disks.
Since direct observations and measurements are hard to perform for many of these flows, laboratory models that incorporate the essential features of these flows can be useful.
A common simple rotating shear flow that can be implemented in the laboratory is Taylor-Couette (TC) flow, which is the flow in the fluid-filled gap between two coaxial rotating cylinders.
Taylor-Couette flow has found particular applicability as a model for astrophysical accretion disks in determining their stability properties and the outward angular momentum flux which is necessary in order for material to be transported inward onto the central body \citep{zeldovich_ProcRoySoc_1981,richard_zahn_AA_1999,richard_thesis_2001,dubrulle_etal_POF_2005,ji_balbus_PhysToday_2013}.
Taylor-Couette experiments have produced contradictory answers to these questions, causing great debate centered on the effects of the no-slip axial boundaries found in Taylor-Couette experiments which do not match the open stratified boundaries of accretion disks \cite{balbus_nature_2011,avila_PRL_2012,schartman_etal_AA_2012,ji_balbus_PhysToday_2013,edlund_ji_PRE_2014}.

We can define a Reynolds number for the inner (outer) cylinder using the radius $r_i$ ($r_o$), the rotation rate $\Omega_i$ ($\Omega_o$), and the fluid's kinematic viscosity $\nu$, giving

\begin{equation}
	Re_i = \frac{\Omega_i \, r_i \left(r_o - r_i \right)}{\nu}, \quad Re_o = \frac{\Omega_o \, r_o \left(r_o - r_i \right)}{\nu} \; .
	\label{eqn:Rei_Reo}
\end{equation}

Rather than using $Re_i$ and $Re_o$, we use the shear Reynolds number $Re_S$ and the so so-called $q$ parameter, detailed below, to compare different parts of the parameter space.
They have a more intuitive relation to the shear and the global rotation.
The shear Reynolds number $Re_S \propto \left| \Omega_i - \Omega_o \right|$, which quantifies shear, is defined as

\begin{equation}
	Re_S = \frac{2}{1+\eta} \left| Re_i - \eta Re_o \right| \; ,
	\label{eqn:ReS}	
\end{equation}

\noindent where $\eta = r_i / r_o$ is the radius ratio \citep{dubrulle_etal_POF_2005}.
Next to $\eta$, another important geometric quantity is the aspect-ratio $\Gamma = L / \left(r_o - r_i \right)$, which is the ratio of the height of the cylinders $L$ to the gap width.
To quantify the global rotation, we use the $q$ parameter \citep{ji_etal_nature_2006,schartman_etal_AA_2012} defined through the relation

\begin{equation}
	\frac{\Omega_i}{\Omega_o} = \eta^{-q} \; .
	\label{eqn:q}
\end{equation}

\noindent The $q$ parameter is real for co-rotating cylinders, the case exclusively dealt with in this paper.
Hence, we will define both $\Omega_i$ and $\Omega_o$ to be both positive throughout this paper.
Solid-body rotation $\Omega_i = \Omega_o$ corresponds to $q = 0$, $\Omega_i > \Omega_o$ gives $q > 0$, $\Omega_i < \Omega_o$ gives $q < 0$, and pure inner and pure outer rotation correspond to $q = +\infty$ and $q = -\infty$ respectively.

Different dimensionless parameters other than $Re_S$ and $q$ have been used, which are presented here for ease of comparison.
Rather than using $Re_S$ to quantify the shear, previous work on our apparatus \citep{gils_etal_PRL_2011,gils_etal_JFM_2012} has used a Taylor number $Ta = \left(\sigma \, Re_S \right)^2$ where $\sigma = \left(1+\eta \right)^4 / \left(2 \sqrt{\eta}\right)^4$ is a geometric Prandtl number \citep*{eckhardt_etal_JFM_2007}, which equals $1.057$ for our $\eta = 0.716$.
With a nearly identical geometry, \citet{paoletti_lathrop_PRL_2011} used a different definition of the Reynolds number, namely $Re = \sqrt{\sigma} \, Re_S = \sqrt{Ta / \sigma}$.
Another parameter quantifying global rotation is the rotation parameter, $R_\Omega \,$ \citep{dubrulle_etal_POF_2005} defined as $R_\Omega = \left(1 - \eta \right) \left(Re_i + Re_o \right) / \left(\eta Re_o - Re_i \right)$.

At low $Re_S$, before the formation of Taylor-vortices, and in the absence of Ekman pumping from axial boundaries (e.g. periodic or free-slip axial boundary conditions); the azimuthal velocity profile is

\begin{equation}
	u_{\theta , \mathrm{lam}}(r) = A r + \frac{B}{r}, \quad A = \frac{\Omega_o - \eta^2 \Omega_i}{1 - \eta^2}, \quad B = \frac{r_i^2 \left( \Omega_i - \Omega_o \right)}{1 - \eta^2} \; .
	\label{eqn:laminar_couette}
\end{equation}

\noindent We will refer to this as laminar Taylor-Couette flow.

\begin{figure}
	\centerline{\includegraphics{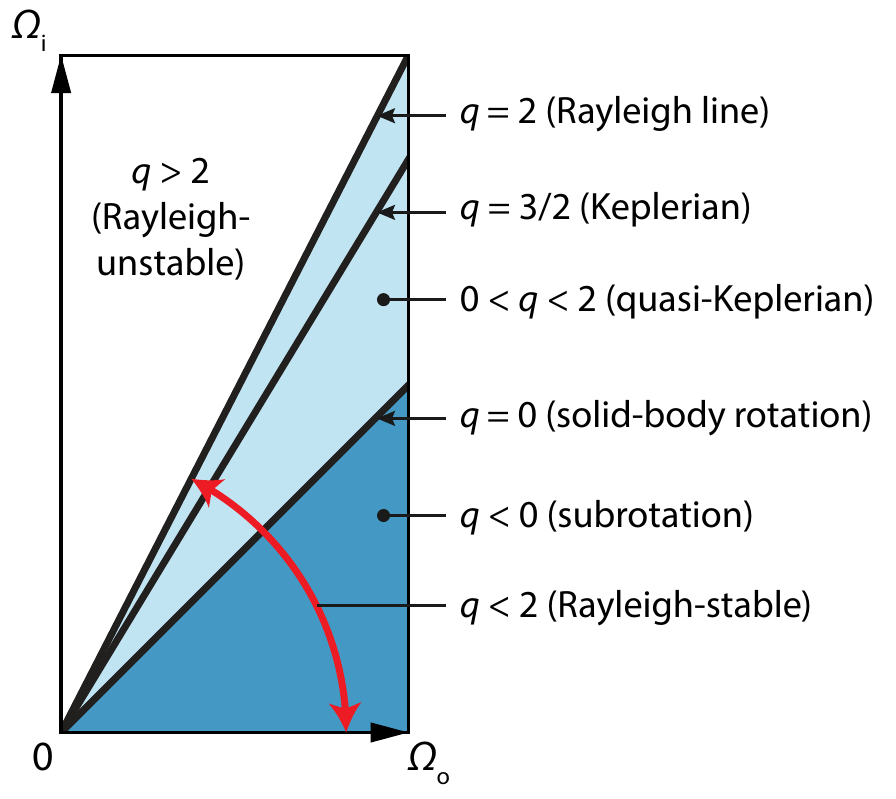}}
	\caption{Taylor-Couette parameter space considered here (red arrow).  The different regions are shown as well as three important lines of constant $q$: the Rayleigh line ($q = 2$), the Keplerian line ($q = 3/2$), and solid-body rotation ($q = 0$).  The whole Rayleigh-stable region ($q < 2$) is shaded with different colors for the quasi-Keplerian regime (light blue) and the sub-rotating regime (dark blue).}
	\label{fig:parameter_space}
\end{figure}

Some rotating flows have radially increasing specific angular momentum ($\sign{\left({\partial \ell}/{\partial r}\right)} = \sign{\ell}$), where $\ell = r^2 \omega$ is the specific angular momentum and $\omega = u_\theta / r$ is the fluid angular velocity.
Such flows (see figure~\ref{fig:parameter_space}); as long as they are purely hydrodynamic, barotropic, and stably stratified as we consider here; are stable to infinitesimal perturbations (i.e. linearly stable) according to the Rayleigh criterion \citep{rayleigh_prsa_1917}.
For Taylor-Couette flow, this corresponds to $q < 2$.
Flows for which $q > 2$ are linearly unstable at sufficiently high Reynolds numbers \citep{taylor_prsa_1923}, which is often called the centrifugal instability.
Hence $q = 2$ is referred to as the Rayleigh line.
The Rayleigh-stable region includes sub-rotation ($\Omega_i < \Omega_o$), solid-body rotation ($\Omega_i = \Omega_o$), and super-rotation ($\Omega_i > \Omega_o$).
The flow in the super-rotating region is often referred to as quasi-Keplerian, since it includes cylinder rotation rates ($q = 3/2$) obeying Kepler's 3rd law relating orbital radius and period.
This regime is of particular relevance to astrophysical systems such as accretion disks since they are Rayleigh-stable with azimuthal flow profiles in the plane of the disk that are expected to not deviate significantly from Kepler's 3rd law when ignoring the disk's self-gravitation and relativistic effects \citep{richard_zahn_AA_1999,richard_thesis_2001,dubrulle_etal_POF_2005,ji_balbus_PhysToday_2013}.

Accretion disks are Rayleigh-stable but are known to have accretion rates requiring radial fluxes of angular momentum far greater than the flux provided by viscous diffusion in laminar Taylor-Couette flow-like disks, indicating that they are in fact unstable \citep{richard_zahn_AA_1999,richard_thesis_2001,dubrulle_etal_POF_2005,ji_balbus_PhysToday_2013}.
There has been a search for the instabilities at play in these flows.
Disks sufficiently ionized to be electrically conductive are known to be unstable via the Magneto Rotational Instability (MRI) \citep[][and description therein]{ji_balbus_PhysToday_2013}.
For weakly ionized disks or parts of disks, investigation has focused on stability in the presence of stratification \citep{dubrulle_etal_AA_2005,bars_gal_PRE_2007,ledizes_riedinger_JFM_2010} and stability to finite amplitude perturbations (non-linear stability) which has been the subject of several Taylor-Couette experiments including our own in this paper \citep{richard_thesis_2001,ji_etal_nature_2006,paoletti_lathrop_PRL_2011,paoletti_etal_AA_2012,schartman_etal_AA_2012,edlund_ji_PRE_2014}.

For an incompressible fluid in the Rayleigh-stable region of Taylor-Couette flow and compressible accretion disk flow, the possibility of a non-linear instability has not yet been ruled out for all $Re_S$.
Plane Couette flow and pipe flow are both examples of linearly stable flows that have non-linear instabilities at sufficient $Re$ \citep[][and references therein]{grossmann_rmp_2000,avila_etal_science_2011,shi_etal_PRL_2013}.
\citet{maretzke_hof_avila_JFM_2014} found transient growth, a necessary prerequisite for a non-linear instability, in Rayleigh-stable Taylor-Couette flow.
Accretion disks have very high Reynolds numbers with $Re_S$ possibly as high as $10^{14}$ \citep{paoletti_etal_AA_2012,ji_balbus_PhysToday_2013}.
Therefore, it is reasonable to ask whether Rayleigh-stable Taylor-Couette flow is non-linearly stable or unstable.

In prior experimental work; visualization via Kalliroscope particles, angular momentum transport measurements, and velocimetry measurements were done; yielding contradictory results on the presence of a non-linear instability, especially for quasi-Keplerian flow \citep{wendt_1933,taylor_ProcRoySoc_1936_a,taylor_ProcRoySoc_1936_b,coles_JFM_1965,richard_thesis_2001,ji_etal_nature_2006,borrero_etal_PRE_2010,paoletti_lathrop_PRL_2011,burin_czarnocki_JFM_2012,schartman_etal_AA_2012,paoletti_etal_AA_2012, edlund_ji_PRE_2014}.
These experiments have, to varying degree, Ekman pumping driven by the no-slip boundary conditions on the axial boundaries.
The Ekman pumping could destabilize the flow depending on the axial end configuration in a way that would not be found in astrophysical accretion disks \citep{balbus_nature_2011,avila_PRL_2012,schartman_etal_AA_2012,ji_balbus_PhysToday_2013, edlund_ji_PRE_2014}, which have open stratified axial boundaries.
Axial boundaries that rotate with the outer cylinder, such as those on the apparatus presented in this paper, were found to have Ekman pumping effects that spanned the whole flow volume \citep{avila_PRL_2012,schartman_etal_AA_2012, edlund_ji_PRE_2014}, which might explain the large, and likely turbulent, angular momentum transport found by the Maryland experiment \citep{paoletti_lathrop_PRL_2011} in contrast to the low angular momentum transport steady laminar flow found in the Princeton MRI and HTX experiments which reduced the Ekman pumping by splitting the axial boundaries into rings rotated at speeds intermediate that of the two cylinders \citep{ji_etal_nature_2006,schartman_etal_AA_2012,edlund_ji_PRE_2014}.

The effect of the Ekman pumping in wide-gap ($\eta < 0.34$) low aspect-ratio ($\Gamma < 3$) Rayleigh-stable experiments, such as the Princeton MRI and HTX experiments, on the flow state and angular momentum transport has been the subject of several investigations.
When the axial boundaries are attached to the outer cylinder as opposed to rotating at intermediate speeds, there are large fluctuations and mixing near the inner cylinder \citep{dunst_JFM_1972,edlund_ji_PRE_2014} and quiescent flow rotating close to $\Omega_o$ near the outer cylinder \citep{dunst_JFM_1972,kageyama_etal_JournPhysSocJapan_2004,schartman_etal_AA_2012,edlund_ji_PRE_2014}.
Speeding up the part of the axial boundaries near the inner cylinder causes the fluctuations near the inner cylinder to decrease and the azimuthal velocities to more closely match laminar Taylor-Couette flow \citep{edlund_ji_PRE_2014}.
In the reduced Ekman pumping configuration, perturbations by jets from the inner cylinder were found to decay for $Re_S \le 10^6$ \citep{edlund_ji_PRE_2014}.

The effects of the Ekman pumping in medium-gap ($\eta \approx 0.7$) larger aspect-ratio ($\Gamma \sim 10$) experiments, such as the Maryland and our experiments, has not received as much attention, though it has been expected to be similar, which would resolve the contradictory results.
At low $Re_S < 10^4$, axial boundaries attached to the outer cylinder were found to destabilize the flow \citep{avila_PRL_2012}.
For $Re_S \le 10^5$ Direct Numerical Simulations (DNS) with periodic axial boundaries in the quasi-Keplerian regime, \citet{monico_etal_JFM_2014b} found that initial turbulent states always decayed to laminar Taylor-Couette flow.
In this paper, we present azimuthal velocimetry profiles in both the quasi-Keplerian and the sub-rotating regimes in a geometry similar to the Maryland experiment.
We compare them to the profiles in laminar Taylor-Couette flow and discuss their structure to better elucidate the effects of Ekman pumping on the flow and the large angular momentum transport associated with axial boundaries fixed to the outer cylinder for our geometry \citep{paoletti_lathrop_PRL_2011}.

The paper is organized as follows: section~\ref{sec:experiment} describes the experiment and the parameter space explored, section~\ref{sec:results} presents the azimuthal velocity and specific angular momentum profiles, section~\ref{sec:analysis} presents further analysis and discussion of the azimuthal profiles including the primarily axial transport of angular momentum, and section~\ref{sec:conclusions} summarizes the results and presents conclusions.

\section{Experiment and explored parameter space} \label{sec:experiment}

\begin{figure}
	\centerline{\includegraphics[height=8.5cm]{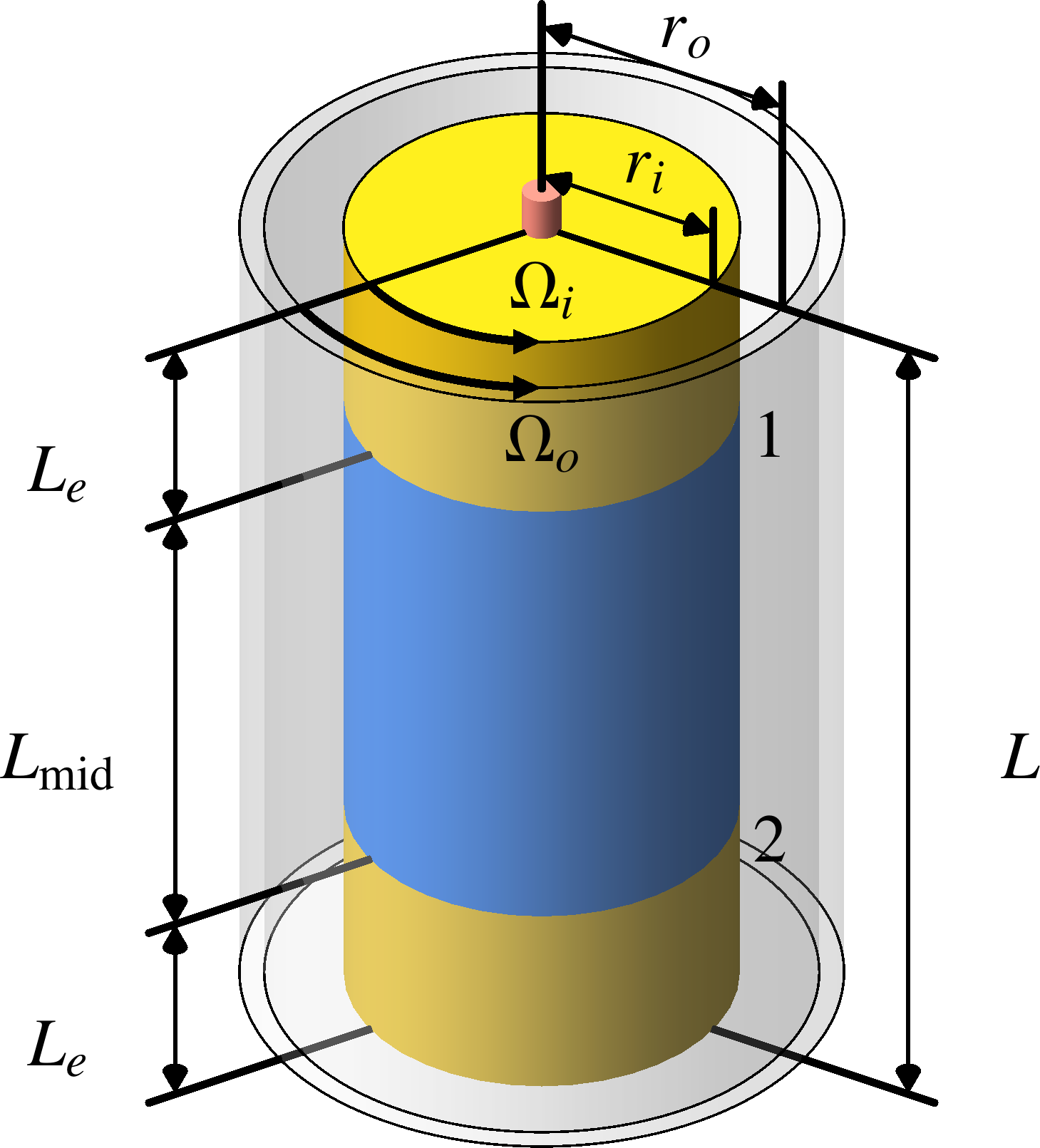}}
	\caption{Sketch of the T$^3$C apparatus used for the measurements presented in this paper.  The inner cylinder is split into three sections, having a $2.5$~mm gap between each section marked as positions 1 and 2.}
	\label{fig:apparatus}
\end{figure}

The apparatus is described in detail in \citet{gils_etal_RevSciInst_2011} and is shown schematically in figure~\ref{fig:apparatus}.
A brief summary is given here.
The inner cylinder has an outer radius of $r_i = 20.00$~cm and the transparent outer cylinder has an inner radius of $r_o = 27.94$~cm, which gives $\eta = 0.716$.
They can independently rotate up to a maximum of  $\left| \Omega_i / 2 \pi \right| = 20$~Hz and $\left| \Omega_o / 2 \pi \right| = 10$~Hz, respectively.
The total height is $L = 93.2$~cm, which gives $\Gamma = 11.74$.
The axial boundaries are attached to the outer cylinder.
The inner cylinder is split into three sections.
The height of the middle section is $L_\mathrm{mid} = 53.6$~cm, and the end sections have equal heights of $L_e = 19.35$~cm.
There is a $2.5$~mm gap between each section, labeled 1 and 2 in figure~\ref{fig:apparatus}.
The system was filled with water and operated at room temperature with cooling applied at the axial boundaries.
This geometry is similar to the apparatus used by \citet{paoletti_lathrop_PRL_2011}, who have a $\eta = 0.7245$ and $\Gamma = 11.47$ geometry.

\begin{table}
	\centering
	\begin{tabular}{ccclccc}
		 & & & & & $Re_S$ & \\
		\cline{5-7} \\
		$q$ & $\;\; \Omega_i / \Omega_o \;\;$ & $R_\Omega$ & Region & $\; 2.07 \times 10^4 \;$ & $\; 1.04 \times 10^5 \;$ & $\; 7.81 \times 10^5 \;$ \\
		\hline
			$\;\;\;2.100$ &	$2.018\;\;$ & $-0.9533$ & Rayleigh unstable & & \checkmark & \checkmark \\
			$\;\;\;1.909$ &	$1.893\;\;$ & $-1.047\;\;$ & quasi-Keplerian & & \checkmark & \checkmark \\
			$\;\;\;1.500$ &	$1.651\;\;$ & $-1.333\;\;$ & quasi-Keplerian & \checkmark & \checkmark & \checkmark  \\
			$\;\;\;1.258$ &	$1.523\;\;$ & $-1.587\;\;$ & quasi-Keplerian & & \checkmark & \\
			$\;\;\;0.692$ &	$1.260\;\;$ & $-2.900\;\;$ & quasi-Keplerian & & \checkmark & \\
			$\;\;\;0.333$ &	$1.118\;\;$ & $-6.062\;\;$ & quasi-Keplerian & & \checkmark & \\
			$-0.500$ &	$0.8461$ & $\;\;\;4.141\;\;$ & sub-rotating & & \checkmark & \\
			$-1.000$ &	$0.7158$ & $\;\;\;2.113\;\;$ & sub-rotating & & \checkmark & \\
			$-2.000$ &	$0.5124$ & $\;\;\;1.113\;\;$ & sub-rotating & & \checkmark & \\
	\end{tabular}
	\caption{The $q$ values for which velocity profiles were measured and their corresponding rotation rate ratios ($\Omega_i / \Omega_o$) and rotation parameter, $R_\Omega$.  We also give the region of the Taylor-Couette parameter space the measurement is in, and for what $Re_S$ measurements were taken.}
	\label{table:qvalues}
\end{table}

The azimuthal velocity profiles were obtained using Laser Doppler Anemometry (LDA).
The LDA configuration used backscatter from seed particles in a measurement volume of approximately $0.07$~mm~$\times$~$0.07$~mm~$\times$~$0.3$~mm.
Dantec PSP{-}5 particles with a $5$~$\mu$m diameter and $1.03$~$\mathrm{g} / \mathrm{cm}^3$ density were used.
The optical effect of the outer cylinder curvature on the LDA measurements was corrected by using the calculations of \citet{huisman_etal_EJMBF_2012}.
The velocimetry was calibrated using radial and axial profiles of solid-body rotation at different rotation rates.
The error in the mean velocity profiles from the calibration, which was the dominant source in the mean profiles, was smaller than $0.1\%$.
For all LDA measurements a statistical convergence of $1\%$ was achieved, which translates to between $1\%$ and $6\%$ of $\left| \Omega_i - \Omega_o \right|$, which prevents investigation into fluctuations and deviations from axisymmetry.
When measuring close to the inner cylinder, reflections from the metal inner cylinder were found to be problematic.
Hence, the radial profiles presented in this paper were done at the axial height of the $2.5$~mm gap between the bottom and middle inner cylinder sections, which corresponds to an axial height $z / L = 0.209$ off the bottom, so that the LDA laser would be absorbed in the gap as opposed to being reflected off the cylinder surface.
The axial dependence of the angular velocity was found to be less than $2\%$ of $\left| \Omega_i - \Omega_o \right|$ from axial profiles at midgap from midheight to $1.5$~cm off the bottom, and between radial profiles over the outer half of the gap at five heights $z = \left\{ 0.195, \, 0.223, \, 0.414, \, 0.464, \, 0.927 \right\}$~m off the bottom, which are at $z/L = \left\{ 0.209, \, 0.238, \, 0.444, \, 0.497, \, 0.995 \right\}$.
The last one, $z/L = 0.995$ is 5~mm from the top axial boundary.
Thus, a radial profile at $z / L = 0.209$ is representative, other than possibly for radial positions closer than $2.5$~mm to the inner cylinder.
The boundary layers on the axial boundaries are confined to within 5~mm of the boundaries.

Velocimetry was performed for five quasi-Keplerian $q$ values including Keplerian ($q = 1.500$), three sub-rotating values of $q$, and one unstable but very close to the Rayleigh line $q$ value ($q = 2.100$); which are all listed in table~\ref{table:qvalues}.
The value $q = 1.909$ was chosen to match the simulations of \citet{avila_PRL_2012} on a nearly identical geometry and the Princeton experimental work at $q = 1.9$ \citep{ji_etal_nature_2006,schartman_etal_AA_2012}.
Also, $q = \left\{ 1.909, \, 1.500, \, 1.258, \, 0.692 \right\}$ were chosen to match ongoing torque measurements on the Maryland experiment.
Measurements for all values of $q$ were taken at $Re_S = 1.04 \times 10^5$, the three values of $q \ge 1.500$ at $Re_S = 7.81 \times 10^5$, and $q = 1.500$ at $Re_S = 2.07 \times 10^4$.
All of the azimuthal velocity profiles, radial profiles at all 5 heights and the axial profile at midgap, are available in the supplementary material.
Each pair of $q$ and $Re_S$ was reached by starting with both cylinders at rest, linearly increasing $\Omega_i$ and $\Omega_o$ to their final values over $120$~s while maintaining constant $q$, and then waiting at least $600$~s for transients to decay before doing measurements.

\section{Results on the azimuthal profiles} \label{sec:results}

It is convenient to look at the velocity profiles in terms of the normalized radial position and the normalized angular velocity given by

\begin{eqnarray}
	\widetilde{r} & = & \frac{r - r_i}{d} \label{eqn:rnorm} \; , \\
	\widetilde{\omega} & = & \frac{\omega - \Omega_o}{\Omega_i - \Omega_o} \; , \label{eqn:wnorm}
\end{eqnarray}

\noindent where $d = r_o - r_i$ is the width of the gap.
The expression for $\widetilde{r}$ gives $\widetilde{r} = 0$ at the inner cylinder and $\widetilde{r} = 1$ at the outer cylinder.
Regardless of which cylinder has the larger angular velocity, the expression for $\widetilde{\omega}$ gives $\widetilde{\omega} = 0$ whenever $\omega = \Omega_o$ and $\widetilde{\omega} = 1$ whenever $\omega = \Omega_i$.
For the quasi-Keplerian regime, $\widetilde{\omega} > 1$ indicates a super-rotating flow with $\omega > \Omega_i > \Omega_o > 0$, and $\widetilde{\omega} < 0$ indicates a sub-rotating flow with $\Omega_i > \Omega_o > \omega$.
Due to the sign change in the denominator for the sub-rotating regime, $\widetilde{\omega} > 1$ indicates a sub-rotating flow with $\Omega_o > \Omega_i > \omega$ and $\widetilde{\omega} < 0$ implies a super-rotating flow with $\omega > \Omega_o > \Omega_i > 0$.
The laminar Taylor-Couette profile, which in these normalized variables is independent of $\Omega_i$ and $\Omega_o$, is

\begin{equation} 
	\widetilde{\omega}_\mathrm{lam} = \frac{\eta^2 \left( 1 - \widetilde{r} \right) \left( \widetilde{r} \left( 1 - \eta \right) + 1 + \eta \right)}{\left( 1 + \eta \right) \left( \widetilde{r} \left( 1 - \eta \right) + \eta \right)^2} \; .
	\label{eqn:wnorm_lam}
\end{equation}

\begin{figure}
	\centerline{
		\subfloat[][]{\label{fig:profiles_qDependence_ReS=1e5_full} \includegraphics{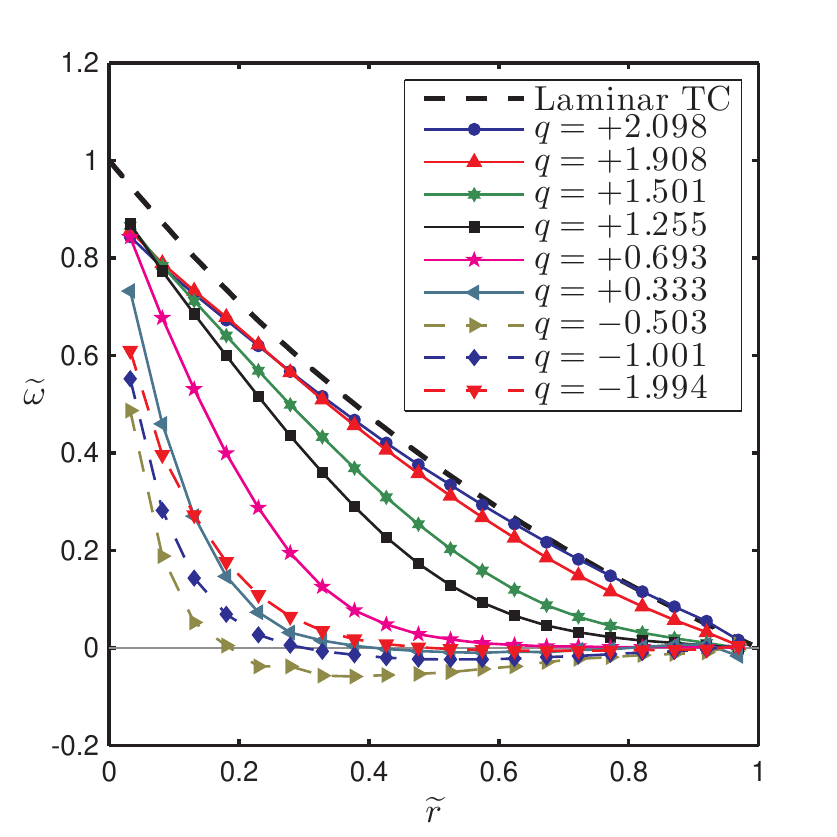}}
		\hspace{-0.8cm}
		\subfloat[][]{\label{fig:profiles_qDependence_ReS=1e5_zoomed} \includegraphics{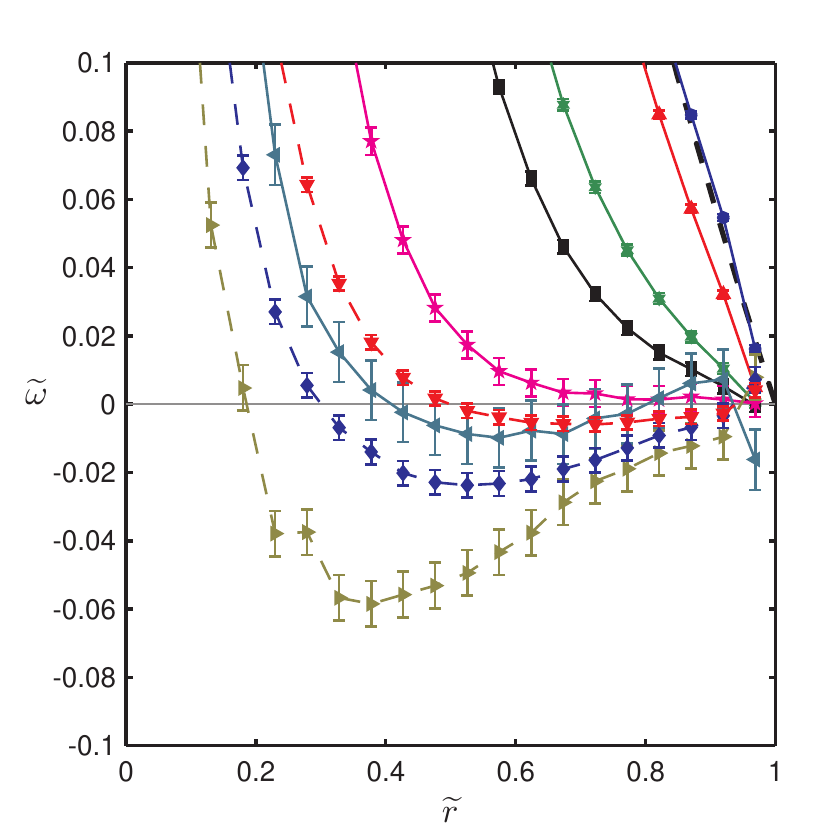}}
	}
	\caption{Comparison of the normalized angular velocity $\widetilde{\omega} =  \left(\omega - \Omega_o \right) / \left( \Omega_i - \Omega_o \right)$ profiles across the gap for different values of the $q$ parameter at $Re_S = 1.04 \times 10^5$.  \protect\subref{fig:profiles_qDependence_ReS=1e5_full} shows the full scale of $\widetilde{\omega}$ (error bars are smaller than the symbols) and  \protect\subref{fig:profiles_qDependence_ReS=1e5_zoomed} shows an expansion around $\widetilde{\omega} = 0$, using the same symbols to emphasize the parts of the profiles close to rotation at $\Omega_o$.  Connecting lines are drawn to guide the eye.  The profile for laminar Taylor-Couette flow is drawn for comparison.}
	\label{fig:profiles_qDependence_ReS=1e5}
\end{figure}

\begin{figure}
	\centerline{\includegraphics{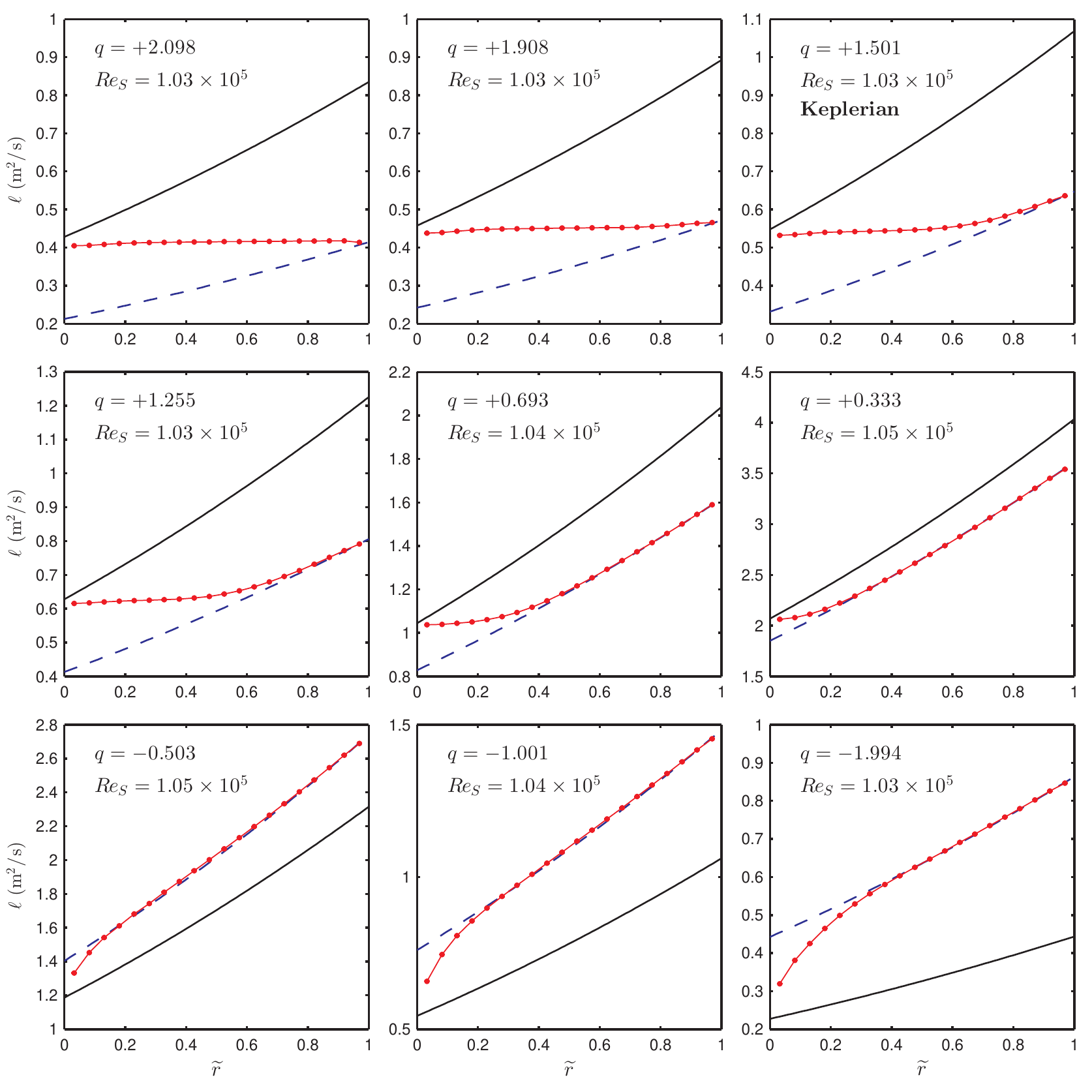}}
	\caption{The specific angular momentum ($\ell = r^2 \omega$) profiles across the gap for the different $q$ values at $Re_S \approx 1.04 \times 10^5$.  The red circles (\textcolor{red}{$\bullet$}) are the specific angular momentum profiles of the flow, with connecting lines to guide the eye.  Error bars are smaller than the symbols.  The solid black line ($\pmb{\bm{-}}$) and dashed blue line ($\textcolor{blue}{\pmb{\bm{--}}}$) are the specific angular momentum profiles for $\omega\left(\widetilde{r}\right) = \Omega_i$ and $\omega\left(\widetilde{r}\right) = \Omega_o$, respectively. The vertical axes have the same units and the horizontal axes are the same for all plots.  The Keplerian configuration is shown at the top-right.}
	\label{fig:angularmomentumProfiles}
\end{figure}

\begin{figure}
	\centerline{
		\subfloat[][]{\label{fig:normalizedAngularVelocityProfiles_ReSDependence_Keplerian} \includegraphics{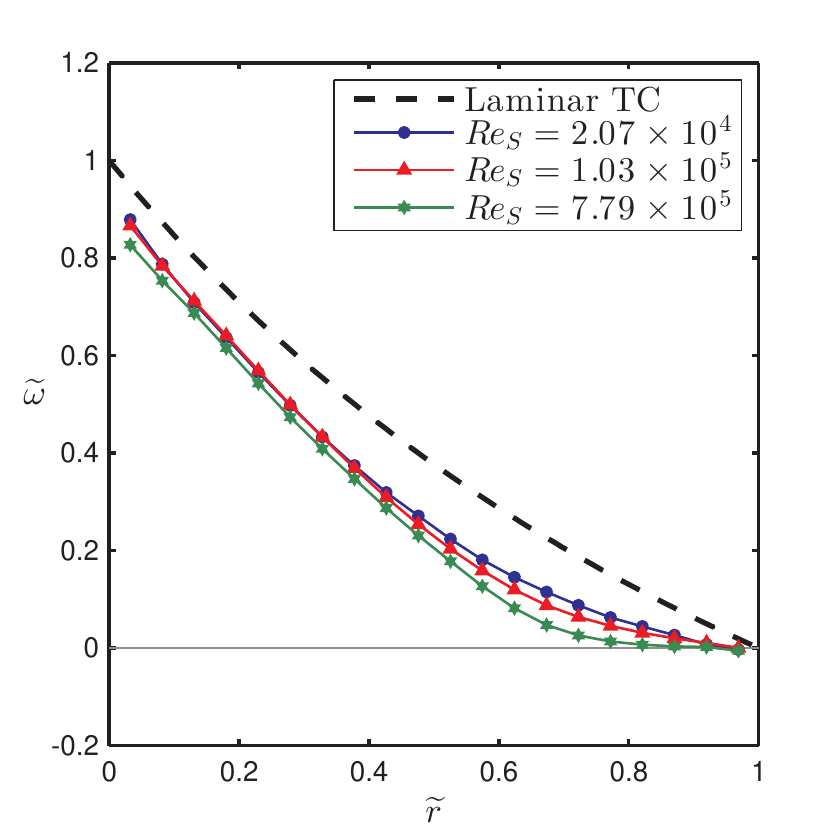}}
		\hspace{-0.8cm}
		\subfloat[][]{\label{fig:angularMomentumProfiles_ReSDependence_Keplerian} \includegraphics{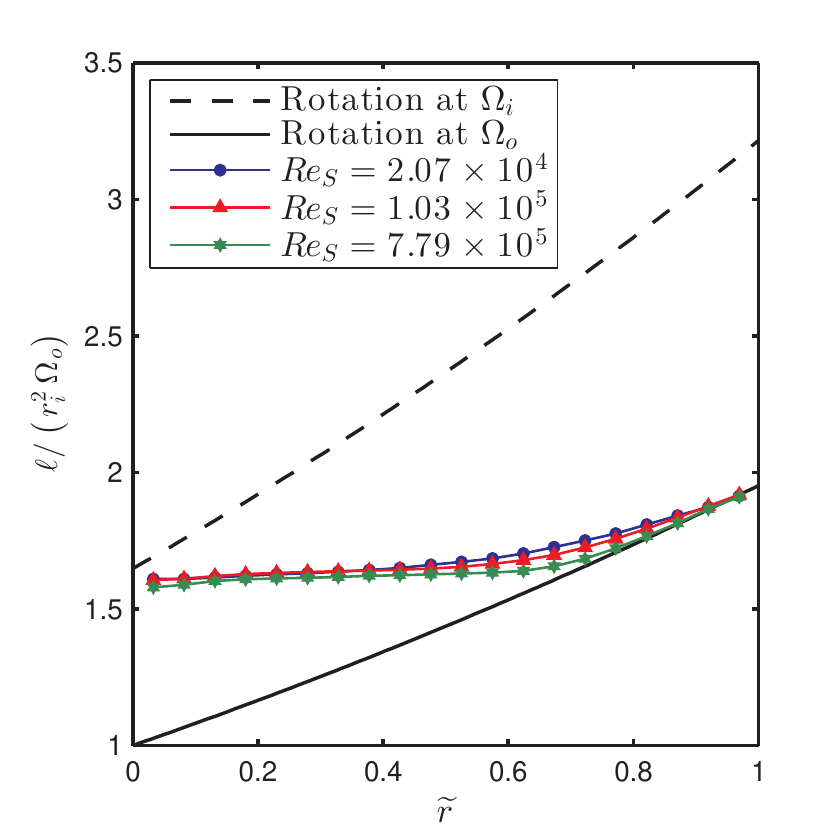}}
	}
	\caption{Comparison of the normalized angular velocity and specific angular momentum profiles across the gap for the Keplerian cylinder rotation ratio ($q = +1.500$) at three different $Re_S$.  \protect\subref{fig:normalizedAngularVelocityProfiles_ReSDependence_Keplerian} the normalized angular velocity $\widetilde{\omega}$ with the laminar Taylor-Couette profile drawn for comparison and \protect\subref{fig:angularMomentumProfiles_ReSDependence_Keplerian} the specific angular momentum $\ell$ normalized by $r_i^2 \, \Omega_o$ with lines for solid-body rotation at the inner and outer cylinder rotation rates ($r^2 \Omega_i$ and $r^2 \Omega_o$, respectively).  The error bars are smaller than the symbol heights.}
	\label{fig:profiles_ReSDependence_Keplerian}
\end{figure}

The $\widetilde{\omega}$ profiles for all values of $q$ at $Re_S = 1.04 \times 10^5$ are compared to each other and to the laminar Taylor-Couette profile in figure~\ref{fig:profiles_qDependence_ReS=1e5}.
None of the profiles matched the laminar Taylor-Couette profile.
Approaching solid-body rotation ($q \rightarrow 0$) at fixed $Re_S$ in both regimes, deviation from the laminar Taylor-Couette profile increased and the part of the profile near the inner cylinder steepened.
For the quasi-Keplerian regime, as we approach solid-body rotation, the rest of the profile flattens towards $\widetilde{\omega} = 0$.
For the sub-rotating regime, $\widetilde{\omega} < 0$ away from the inner cylinder.
This indicates that the fluid is \emph{super-rotating} in terms of angular velocity compared to both cylinders ($\omega > \Omega_o > \Omega_i > 0$) with the degree of super-rotation, as a fraction of $\left| \Omega_i - \Omega_o \right|$, increasing as we approach solid-body rotation.
This flow super-rotation will be further discussed in section~\ref{sec:superrotation}.

The resulting profiles of the specific angular momentum $\ell = r^2 \omega$ at $Re_S = 1.04 \times 10^5$ are shown in figure~\ref{fig:angularmomentumProfiles}.
For the quasi-Keplerian regime, the specific angular momentum profiles all follow the same pattern of having an inner flat region connected to an outer region rotating at $\Omega_o$, which will be discussed further in section~\ref{sec:quasiKeplerian}.
The flat region in $\ell$ indicates that the flow was well mixed in that region.

Keplerian ($q = +1.500$) profiles for three different $Re_S$ are compared in figure~\ref{fig:profiles_ReSDependence_Keplerian}.
They all have a similar shape; but as $Re_S$ is increased, $\widetilde{\omega}$ decreases towards solid-body rotation at $\Omega_o$, especially in the outer parts of the gap.
In terms of the specific angular momentum, increasing $Re_S$ leads to a sharper transition between the flat region and the rotation at $\Omega_o$ region.

\section{Further analysis and discussion} \label{sec:analysis}

\subsection{Super-rotating flow for the sub-rotating regime} \label{sec:superrotation}

\begin{figure}
	\centerline{
		\subfloat[][]{\label{fig:superrotation_levels} \includegraphics{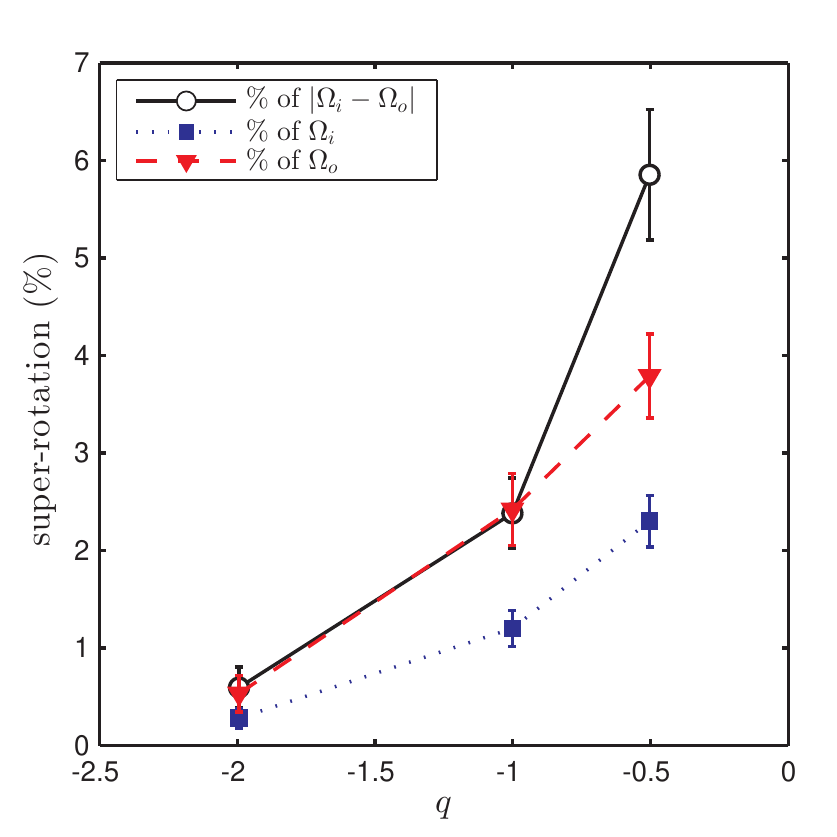}}
		\hspace{-0.4cm}
		\subfloat[][]{\label{fig:region_boundaries_subrotating} \includegraphics{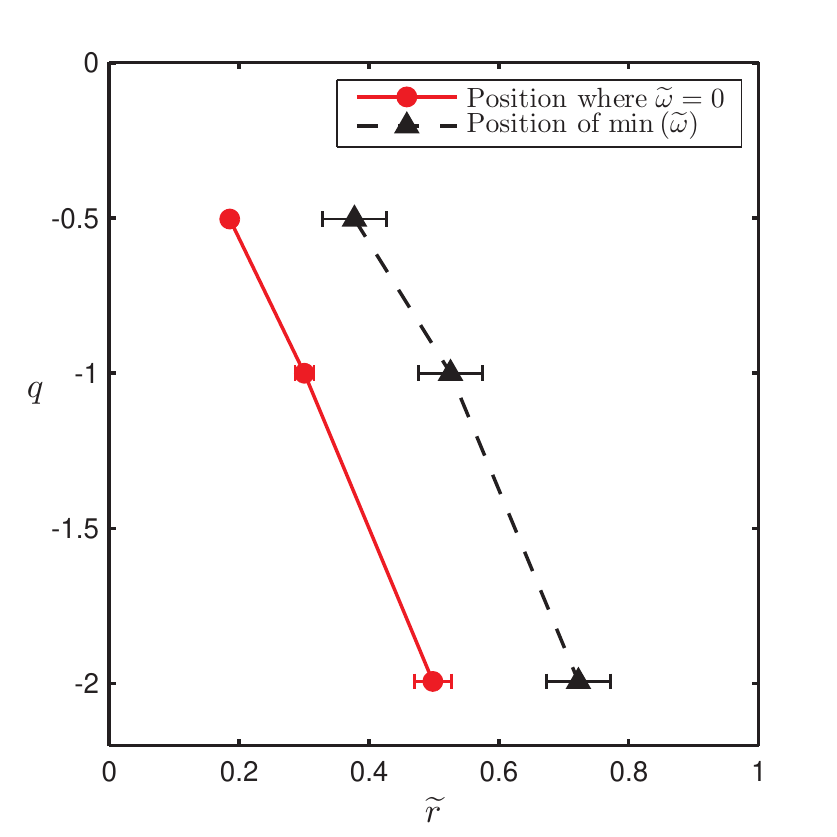}}
	}
	\caption{Super-rotating flow strength and locations in the sub-rotating regime at $Re_S = 1.04 \times 10^5$.  \protect\subref{fig:superrotation_levels} Flow super-rotation ($\omega > \Omega_o > \Omega_i > 0$) strength $\omega - \Omega_o$ at each $q$ as a percentage of $\left|\Omega_i - \Omega_o\right|$, $\Omega_i$, and $\Omega_o$. \protect\subref{fig:region_boundaries_subrotating} The radial positions where the profile crosses $\widetilde{\omega} = 0$ ($\omega = \Omega_o$) to be super-rotating, and the radial position where the super-rotation is at its maximum (minimum $\widetilde{\omega}$).  Connecting lines are drawn to guide the eye in both plots.}
	\label{fig:subrotating_quantified}
\end{figure}

As seen in figure~\ref{fig:profiles_qDependence_ReS=1e5} for all three sub-rotating profiles, $\widetilde{\omega} < 0$ except near the inner cylinder indicating flow super-rotation (figure~\ref{fig:profiles_qDependence_ReS=1e5_zoomed}).
The flow super-rotation can be quantified by taking the minimum $\widetilde{\omega}$ in the profile to be the strength of the super-rotation, and finding its radial position along with where the linear interpolation of where the profile crosses $\omega = \Omega_o$ ($\widetilde{\omega} = 0$) to super-rotation.
The strength of the super-rotation is shown in figure~\ref{fig:superrotation_levels}, and the radial locations of the maximum super-rotation and of $\widetilde{\omega} = 0$ are shown in figure~\ref{fig:region_boundaries_subrotating}.
Approaching solid-body rotation ($q \rightarrow 0$) at fixed non-zero $Re_S$, the strength of super-rotation increases, and the radial positions of the super-rotation maximum and of $\widetilde{\omega} = 0$ both move towards the inner cylinder.
This is a singular limit, which is very different from the limit $q \rightarrow 0$ in which case one would get $\omega\left( r \right) = \Omega_i = \Omega_o$.
The distance between these radial positions was approximately the same for all three $q$, namely a value of $0.2$ gap-widths.
The flow super-rotation was seen at all five heights for which radial profiles of the velocity were taken.
They vary from each other by $\Delta \widetilde{\omega} < 0.01$ axially over the outer half of the gap.
The $0.2$ gap-width separation was seen at the other heights for $q = -0.503$, but could not be resolved for $q = \left\{ -1.001, \, -1.994 \right\}$ since the point where $\widetilde{\omega} = 0$ lies in the inner half of the gap.

The specific angular momentum profiles in figure~\ref{fig:angularmomentumProfiles} were slightly greater than for solid-body rotation with the outer cylinder, except close to the inner cylinder, which is another way of saying there is flow super-rotation.
The Navier-Stokes equation does not constrain angular velocities to be bound by $\Omega_i$ and $\Omega_o$ due to its non-linear term, unlike the temperature field in Rayleigh-B{\'e}nard flow, which is contrained between the two plate temperatures as the temperature advection equation is linear.
Even with the super-rotation, we still have $\partial \ell / \partial r > 0$ over the parts of the gap that are resolved; and $\ell$ is bound between the specific angular momenta of the outer cylinder and the axial boundaries at $\widetilde{r} = 0$, which are the locations of the largest and smallest $\ell$ on the axial boundaries, respectively.
Angular momentum is transported to the inner cylinder in this regime since the torque on the inner cylinder is negative \citep{paoletti_lathrop_PRL_2011}.
With inward advection of angular momentum across the gap (there is also the possibility of axial transport), the outer cylinder and axial boundaries must be the source of angular momentum to sustain the flow super-rotation against spin down to $\omega = \Omega_o$.
This also allows one to estimate the maximum flow super-rotation that could be seen.
If fluid from the outer cylinder having specific angular momentum $\ell = r_o^2 \Omega_o$ is transported to the inner cylinder while conserving $\ell$, it will have an angular velocity $\omega_s = \Omega_o / \eta^2$.
Normalizing the flow super-rotation $\omega_s - \Omega_o$ respectively by $\left|\Omega_i - \Omega_o\right|$, $\Omega_i$, and $\Omega_o$, we get

\begin{eqnarray}
	\frac{\omega_s - \Omega_o}{\left|\Omega_i - \Omega_o\right|} & \le & \left(\frac{1 - \eta^2}{\eta^2}\right) \left|\frac{1}{\eta^{-q} - 1}\right| \; , \label{eqn:superrotation_bound_diff} \\
	\frac{\omega_s - \Omega_o}{\Omega_i} & \le & \left(\frac{1 - \eta^2}{\eta^2}\right) \eta^q \; , \label{eqn:superrotation_bound_inner} \\
	\frac{\omega_s - \Omega_o}{\Omega_o} & \le & \frac{1 - \eta^2}{\eta^2} \; \label{eqn:superrotation_bound_outer}
\end{eqnarray}

\noindent as estimates of the super-rotation upper bound.
For our $\eta = 0.716$, $\left( 1 - \eta^2 \right) / \eta^2 = 0.95$.
The flow super-rotations we see in figure~\ref{fig:superrotation_levels} are one to two orders of magnitude smaller than the estimated bounds.
As equation~(\ref{eqn:superrotation_bound_diff}) diverges as $q \rightarrow 0$, an open question is whether the magnitude of the flow super-rotation normalized by $\left|\Omega_i - \Omega_o\right|$ diverges as $q \rightarrow 0$ at fixed non-zero $Re_S$.

\subsection{Quasi-Keplerian angular momentum profile and transport} \label{sec:quasiKeplerian}

For all the quasi-Keplerian profiles in figure~\ref{fig:angularmomentumProfiles}, there is a pattern in the profiles.
Namely, they are split into three regions: an inner region whose angular momentum profile is nearly flat with a slight positive slope, an outer region where the flow is nearly in solid-body rotation at $\Omega_o$, and a middle transition region in which the angular momentum profile curves upward from being flat to solid-body rotation at $\Omega_o$.
At $q = 1.908$, the inner region extends over nearly the whole gap.
As $q$ decreases for fixed $Re_S$, the inner region shrinks until for $q = 0.333$ it is nearly absent, with the outer region having grown to be almost the whole gap.

As seen in figure~\ref{fig:angularMomentumProfiles_ReSDependence_Keplerian}, as $Re_S$ is increased, the inner and outer regions appear to grow while the middle region shrinks.
The same pattern is seen going from $Re_S = 1.03 \times 10^5$ to $Re_S = 7.82 \times 10^5$ for $q = 1.909$, which is not shown here but can be seen in the data in the supplementary material.
The pattern suggests that in the limit $Re_S \to \infty$, the middle region might disappear entirely.
If we approximate the inner region as a completely flat angular momentum profile, approximate the outer region as rotating at exactly $\Omega_o$, ignore any boundary layer on the inner cylinder, and assume that the pattern holds for the rest of the quasi-Keplerian regime and that no flow state transitions at higher $Re_S$ break it; then the angular velocity profile for the quasi-Keplerian in the asymptotic limit $Re_S \to \infty$ regime in our geometry would be

\begin{eqnarray}
	\omega(r) & = & \left\{ \begin{array}{l l}
    					\left( r_i / r \right)^2 \Omega_i & \quad \text{for $r < r_c$}\\
					& \\
    					\Omega_o & \quad \text{for $r \ge r_c$}
  				\end{array} \right. \label{eqn:asymptotic_profile}
\end{eqnarray}

\noindent with

\begin{eqnarray}
	r_c & = & r_i \sqrt{\frac{\Omega_i}{\Omega_o}} = r_i \, \eta^{-q / 2} \; , \label{eqn:rc} \\
	\widetilde{r}_c & = &  \frac{\eta }{1 - \eta} \left( \eta^{-q/2}-1 \right) \; , \label{eqn:rcnorm} 
\end{eqnarray}

\noindent where $r_c$ is the transition radius between the flat angular momentum profile and solid-body rotation at $\Omega_o$.
For large but finite $Re_S$, equation~(\ref{eqn:asymptotic_profile}) can serve as an approximate profile.
This approximate profile was derived independently by \citet{dunst_JFM_1972} by assuming that the inner region had a flat angular momentum profile, based on his observation of a well-mixed inner region in his Taylor-Couette experiment.

\begin{figure}
	\centerline{\includegraphics{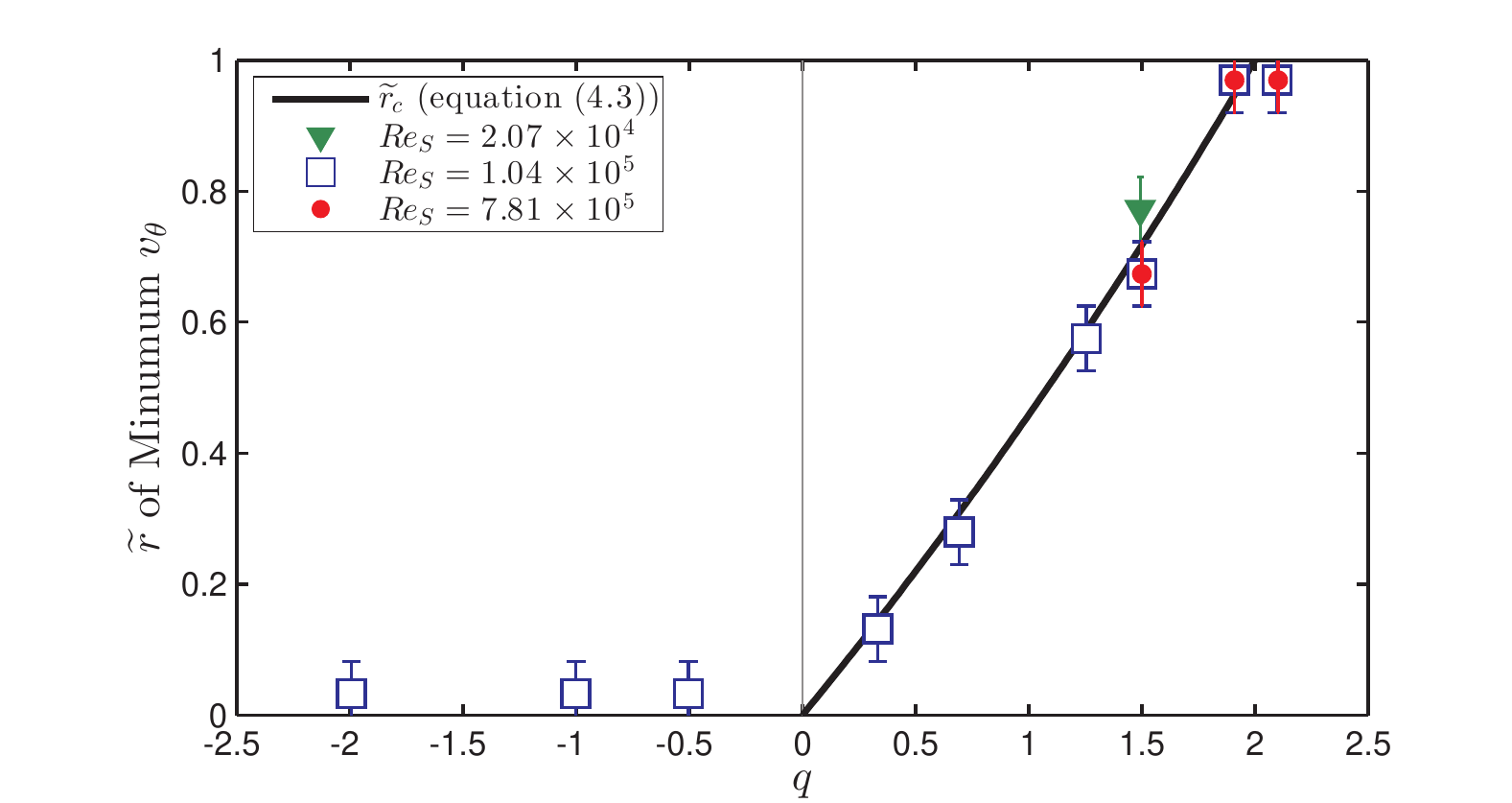}}
	\caption{Radial positions of the minimum in the azimuthal velocity $v_\theta$ as a function of $q$ (x-axis) and $Re_S$ (different symbols).  In the quasi-Keplerian regime, $\widetilde{r}_c$ from equation~(\ref{eqn:rcnorm}) is shown for comparison (solid line).}
	\label{fig:rtrans}
\end{figure}

\begin{figure}
	\centerline{\includegraphics{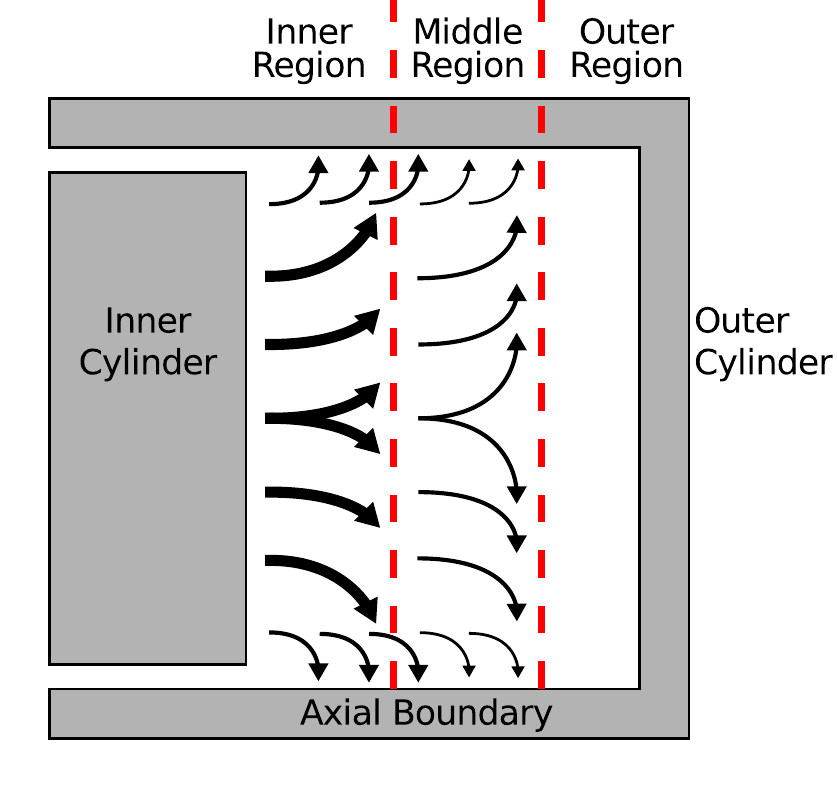}}
	\caption{Schematic drawing of the angular momentum transport in the quasi-Keplerian regime when the axial boundaries are attached to the outer cylinder.  Red dashed lines denote the boundaries between the inner, middle, and outer flow regions.  Black arrows denote the transport of angular momentum.  The radius and aspect-ratios ($\eta$ and $\Gamma$) have been changed for visual clarity.  Angular momentum is transported radially off the inner cylinder and then transported axially to the axial boundaries in the inner and middle regions.}
	\label{fig:angularMomentumTransport_diagram}
\end{figure}

For the approximately constant specific angular momentum $\ell = r^2 \omega = r v_\theta$ inner region, $\partial v_\theta / \partial r < 0$.
Then for the outer regions rotating at approximately $\Omega_i$, $\partial v_\theta / \partial r > 0$.
Hence, we can quantify the radial position of the transition region by finding the radial positions for which the azimuthal velocity profiles $v_\theta(r)$ are at their minimum.
They are shown in figure~\ref{fig:rtrans}.
In the quasi-Keplerian regime, we find that the position of the minimum velocity corresponds very well with $\widetilde{r}_c$ in equation~(\ref{eqn:rcnorm}), giving merit to the approximate profiles of equation~(\ref{eqn:asymptotic_profile}).
Outside of the quasi-Keplerian regime, the position of the minimum is located at the inner cylinder for $q<0$, and at the outer cylinder for $q>2$.

The approximately flat angular momentum profile in the inner region, when away from the Rayleigh line where the laminar Taylor-Couette profile is flat, indicates that the angular momentum is well mixed with advection-dominated transport in the radial direction.
In contrast, there is likely little radial angular momentum transport by advection or diffusion in the outer region as the profile is close to solid-body at $\Omega_o$.
A large amount of angular momentum is transported radially from the inner cylinder based on the torque measurements with the similar Maryland experiment \citep{paoletti_lathrop_PRL_2011} and on the upcoming analysis of section~\ref{sec:torque}.
The large amount of angular momentum transported off the inner cylinder and mixed in the inner region has to go somewhere, but the outer region, if present, is likely not transporting much angular momentum.
Then, when an outer region is present such as when $q < 2$ far from the Rayleigh line, most of the angular momentum must be transported axially to the axial boundaries in the inner and possibly middle regions, as shown schematically in figure~\ref{fig:angularMomentumTransport_diagram}.
As $q$ increases at fixed $Re_S$ towards the Rayleigh line, the outer region disappears and an increasing fraction of the angular momentum can be transported to the outer cylinder through the middle region instead of being transported to the axial boundaries.
For $q \ge 2$, there is no middle region and a boundary layer forms close to the outer cylinder that steepens with increasing $q$ \citep{gils_etal_JFM_2012}, indicating that an increasing fraction of the angular momentum is transported to the outer cylinder instead of to the axial boundaries.
Finally, nearly all of the angular momentum is transported to the outer cylinder.

\begin{figure}
	\centerline{
		\subfloat[][]{\label{fig:compare_profiles_full} \includegraphics{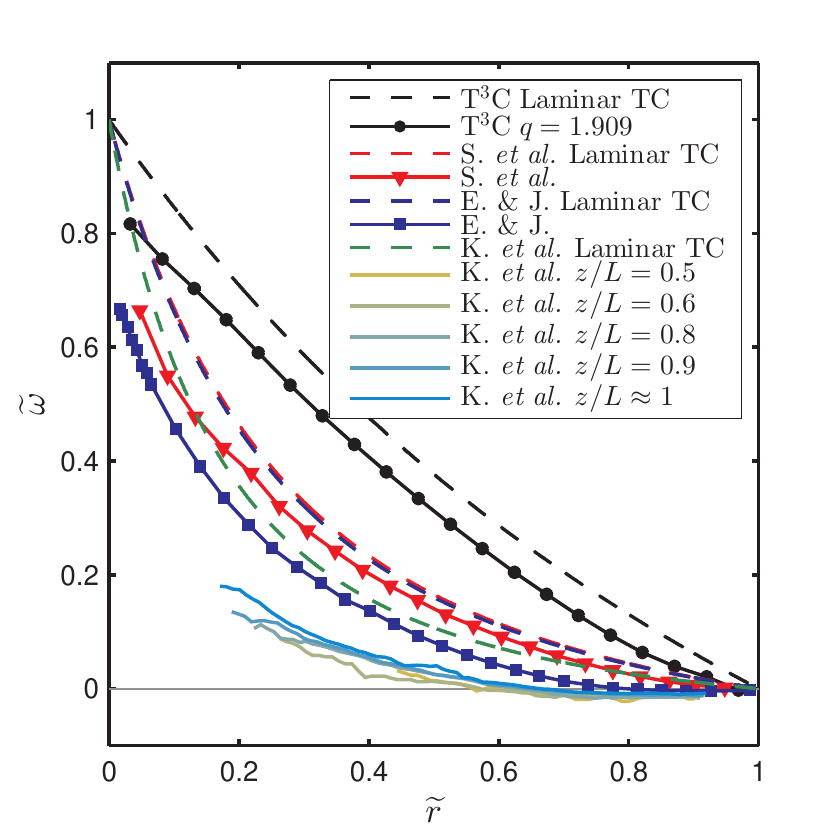}}
		\hspace{-0.8cm}
		\subfloat[][]{\label{fig:compare_profiles_zoomed} \includegraphics{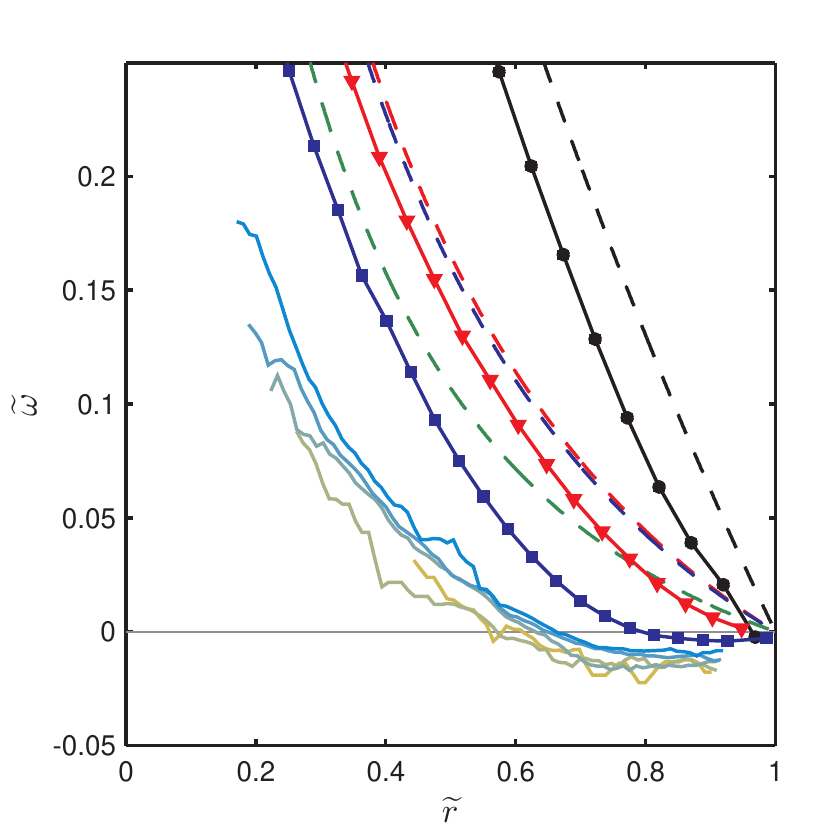}}
	}
	\centerline{
		\subfloat[][]{\label{fig:compare_angular_momentum} \includegraphics{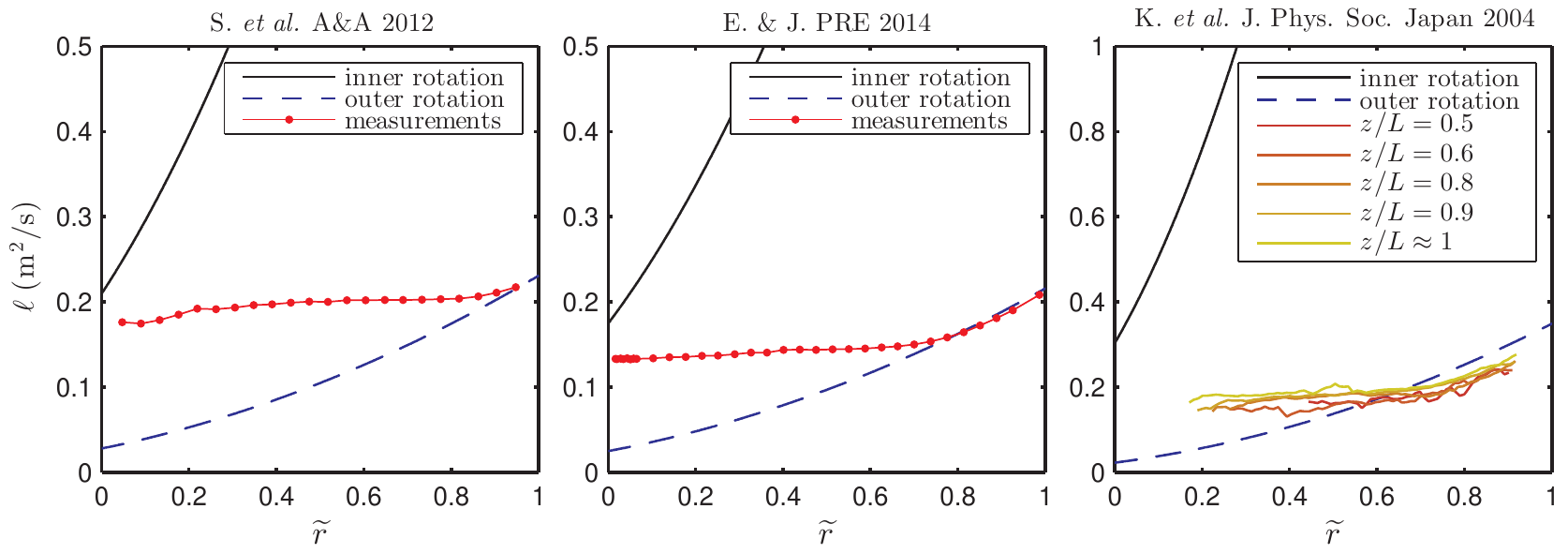}}
	}
	\caption{Comparison of the azimuthal velocimetry for $q \approx 1.9$ between different experiments. This includes our experiment (T$^3$C) with $\eta = 0.716$ at $q = 1.909$ and $Re_S = 7.82 \times 10^5$, \citet{schartman_etal_AA_2012} with $\eta = 0.348$ at $q = 1.908$ and $Re_S = 5.05 \times 10^5$, \citet{edlund_ji_PRE_2014} with $\eta = 0.340$ at $q = 1.803$ and $Re_S = 4.34 \times 10^5$, and \citet{kageyama_etal_JournPhysSocJapan_2004} with $\eta = 0.255$ at $q = 1.896$ and $Re_S = 1.30 \times 10^6$.  The profiles for \citet{schartman_etal_AA_2012} and \citet{edlund_ji_PRE_2014} were constructed by extracting velocities from their figures (6 and 2, respectively).  \citet{kageyama_etal_JournPhysSocJapan_2004} did velocimetry at five different axial heights.  Profiles for \citet{kageyama_etal_JournPhysSocJapan_2004} were constructed by splitting the range $\widetilde{r} \in \left[0, \, 1 \right]$ into bins of width $0.02$ and averaging the $\omega$ within each bin.  Normalized angular velocities $\widetilde{\omega}$ profiles are compared \protect\subref{fig:compare_profiles_full} at full scale and \protect\subref{fig:compare_profiles_zoomed} expanded around $\widetilde{\omega} = 0$ using the same symbols to emphasize the parts of the profiles close to rotation at $\Omega_o$.  Dashed lines are the laminar Taylor-Couette profiles for each experiment.  \protect\subref{fig:compare_angular_momentum} The specific angular momentum ($\ell = r^2 \omega$) profiles for each experiment side by side with the same horizontal axes and with vertical axes in the same units.  The solid black line ($\pmb{\bm{-}}$) and dashed blue line ($\textcolor{blue}{\pmb{\bm{--}}}$) are the specific angular momentum profiles for $\omega\left(\widetilde{r}\right) = \Omega_i$ and $\omega\left(\widetilde{r}\right) = \Omega_o$, respectively.}
	\label{fig:compare_experiments}
\end{figure}

These features are also seen in wide-gap low aspect-ratio experiments.
Using dye injection from the inner cylinder, \citet{dunst_JFM_1972} found a well mixed inner region and a quiescent outer region with poor mixing.
In figure~\ref{fig:compare_experiments}, the angular velocity and the specific angular momentum profiles for $q \approx 1.9$ from \citet{schartman_etal_AA_2012}, \citet{edlund_ji_PRE_2014}, and \citet{kageyama_etal_JournPhysSocJapan_2004} are compared to each other and to the results from our apparatus.
They all deviate from the laminar Taylor-Couette profile and show the same three regions with a relatively flat $\ell$ close to the inner cylinder and rotate close to $\Omega_o$ close to the outer cylinder.
However, the relatively flat $\ell$ inner region is offset downward from the specific angular momentum on the inner cylinder, indicating the presence of a boundary layer on the inner cylinder more significant than in our experiment.
The experiments of \citet{kageyama_etal_JournPhysSocJapan_2004} and possibly \citet{edlund_ji_PRE_2014} also exhibit flow sub-rotation ($\omega < \Omega_o < \Omega_i$) in the middle and outer regions.
The axial transport of angular momentum and the presence of three regions in the quasi-Keplerian azimuthal velocity profiles appear to be more general than just occuring in our specific apparatus with its geometry and ranges of $Re_S$ and $q$, although the strength of the boundary layer on the inner cylinder appears to depend on $\eta$ and/or $\Gamma$.

\subsection{Torque on the inner cylinder} \label{sec:torque}

The velocity gradients near the inner cylinder were larger than in laminar Taylor-Couette as the $\widetilde{\omega}$ values at the point closest to the inner cylinder in figures~\ref{fig:profiles_qDependence_ReS=1e5} and~\ref{fig:normalizedAngularVelocityProfiles_ReSDependence_Keplerian} are below that of the laminar Taylor-Couette profile.
This steepness means that the torque on the inner cylinder must be larger than in laminar Taylor-Couette flow.
If boundary layers were present, the profiles would be even steeper at the inner cylinder, and thus the torques even larger.

The azimuthal shear stress, when averaged azimuthally, is $\tau = - \rho \nu r \left(\partial \omega / \partial r \right)$ where $\rho$ is the fluid density \citep[see page~48,][]{landau_lifshitz_fluids_1987}.
The torque $T$ on a cylinder of radius $r$ from just the shear stress is $T_\nu = 2 \pi r^2 L \tau$, which in terms of the angular velocity is

\begin{equation}
	T_\nu = - 2 \pi \rho \nu L r^3 \frac{\partial \omega}{\partial r} \; .
	\label{eqn:torque_lowerbound}
\end{equation}

As laminar Taylor-Couette flow has no Reynolds stresses and $\omega$ is uniform over a cylinder of radius $r$ from equation~(\ref{eqn:laminar_couette}), the total laminar Taylor-Couette torque is

\begin{equation}
	T_\mathrm{lam} = \frac{2 \pi \rho \nu^2 L \eta}{\left( 1 - \eta \right)^2} Re_S \, \sign{\left( Re_i - \eta Re_o \right)} \; .
	\label{eqn:torque_lam}
\end{equation}

Assuming a turbulent boundary layer, the thickness $y_0$ of the viscous sublayer on the inner cylinder is $y_0 = \left( \nu / u^* \right) y_0^+$ where $u^* = \sqrt{\left| \tau \right| / \rho}$ is the friction velocity, $\rho$ is the fluid density, and $y_0^+$ is the sublayer thickness in dimensionless units \citep{schlichting_boundaryLayerTheory_1979}.
From measurements in our apparatus for pure inner cylinder rotation at comparable $Re_S$, $y_0^+$ is in the range of $5$--$10$ \citep{huisman_etal_PRL_2013}.
Then for $Re_S = 10^5$, we get $y_0 \le 2$~mm since $T \ge T_\mathrm{lam}$ and $y_0^+ \le 10$.
Since $r - r_i = 2.6$~mm was the point closest to the inner cylinder where the flow velocity was resolved, our azimuthal velocimetry did not extend into the viscous sublayer.
Due to not resolving the viscous sublayer, the torque in our apparatus cannot be obtained from the velocity profiles; meaning direct comparisons cannot be done to the torque measurements of \citet{paoletti_lathrop_PRL_2011} on the Maryland experiment with near identical geometry.
However, lower bounds on the torque can be obtained because the azimuthal profiles can give the shear stress, instead of both the shear and Reynolds stresses.

To get the lower bound for the torque on the inner cylinder, $\partial \omega / \partial r$ was obtained from the difference between $\omega$ at the point closest to the inner cylinder ($r - r_i = 2.6$~mm which is $\widetilde{r} = 0.033$) and $\Omega_i$ at the inner cylinder.
It must be noted that the velocity profile was taken at the axial height of one of the small separations in the inner cylinder, which is $2.5$~mm thick, and therefore the gradients in $\omega$ we calculate might be perturbed compared to other axial heights due to the vicinity to the separation.

The torque lower bounds are listed in table~\ref{table:torques}.
The lower bounds were all larger than the laminar Taylor-Couette torque, which supports the $\left| T / T_\mathrm{lam} \right| \gg 1$ result of \citet{paoletti_lathrop_PRL_2011} on the similar Maryland experiment in both regimes for $Re_S > 3.5 \times 10^5$.
The measurements in this paper extend this result of \citet{paoletti_lathrop_PRL_2011} towards solid-body rotation in both regimes.

\begin{table}
	\centering
	\begin{tabular}{ccc}
		$q$ & $\quad \quad \quad Re_S \quad \quad \quad$ & $T_\nu / T_\mathrm{lam}$ \\
		\hline
		$\;\;\;1.493$ &	$2.07 \times 10^4$ &	$2.32 \pm 0.05$ \\
		\\ 
		$\;\;\;2.098$ &	$1.03 \times 10^5$ &	$2.99 \pm 0.04$ \\
		$\;\;\;1.908$ &	$1.03 \times 10^5$ &	$2.75 \pm 0.04$ \\
		$\;\;\;1.501$ &	$1.03 \times 10^5$ &	$2.56 \pm 0.05$ \\
		$\;\;\;1.255$ &	$1.03 \times 10^5$ &	$2.49 \pm 0.06$ \\
		$\;\;\;0.693$ &	$1.04 \times 10^5$ &	$3.02 \pm 0.09$ \\
		$\;\;\;0.333$ &	$1.05 \times 10^5$ &	$5.12 \pm 0.18$ \\
		$-0.503$ &	$1.05 \times 10^5$ &	$9.79 \pm 0.12$ \\
		$-1.001$ &	$1.04 \times 10^5$ &	$8.55 \pm 0.06$ \\
		$-1.994$ &	$1.03 \times 10^5$ &	$7.45 \pm 0.04$ \\
		\\
		$\;\;\;2.102$ &	$7.83 \times 10^5$ &	$3.66 \pm 0.04$ \\
		$\;\;\;1.909$ &	$7.82 \times 10^5$ &	$3.50 \pm 0.04$ \\
		$\;\;\;1.500$ &	$7.79 \times 10^5$ &	$3.31 \pm 0.05$ \\
	\end{tabular}
	\caption{The ratios of the lower bounds of the torque on the inner cylinder $T_\nu$ to the laminar Taylor-Couette torque $T_\mathrm{lam}$ for each set of measurements, ordered by $Re_S$ and then by $q$.}
	\label{table:torques}
\end{table}

For the quasi-Keplerian regime, we can use the approximate flatness of the specific angular momentum profile in the inner region to make an analytical approximate torque lower bound.
Treating the inner region as having a flat specific angular momentum profile from the inner cylinder with no boundary layer as in equation~(\ref{eqn:asymptotic_profile}), the ratio of the torque lower bound $T_{\nu,\mathrm{flat}}$ to the laminar Taylor-Couette torque is

\begin{equation}
	\frac{T_{\nu,\mathrm{flat}}}{T_\mathrm{lam}} = \frac{1 - \eta^2}{1 - \eta^q} \quad \; \mathrm{for} \;\; 0 < q \le 2 \;.
	\label{eqn:torque_lowerbound_quasiKeplerian}
\end{equation}

\noindent The ratio is always larger than one, approaching one at $q = 2$. 
It diverges as $q \to 0$, which is due to the width of the inner region shrinking towards zero since $r_c \to r_i$ in equation~(\ref{eqn:rc}).
The decrease in $r_c$ means that $\omega$ changes from $\Omega_i$ to $\Omega_o$ over an ever smaller radial distance, giving a sharper gradient of $\omega$ in the inner region, which becomes infinite as $q \to 0$.
However, if the inner region of a flat angular momentum profile disappears entirely as $q \to 0$ at a given $Re_S$, then this lower bound may no longer hold.
For $Re_S = 1.04 \times 10^5$, the inner region might be close to disappearing by $q = 0.333$ based on the angular momentum profiles in figure~\ref{fig:angularmomentumProfiles}.
As the middle region shrinks with increasing $Re_S$ (figure~\ref{fig:angularMomentumProfiles_ReSDependence_Keplerian}), the $q$ at which the inner region might disappear decreases with increasing $Re_S$.

\begin{figure}
	\centerline{\includegraphics{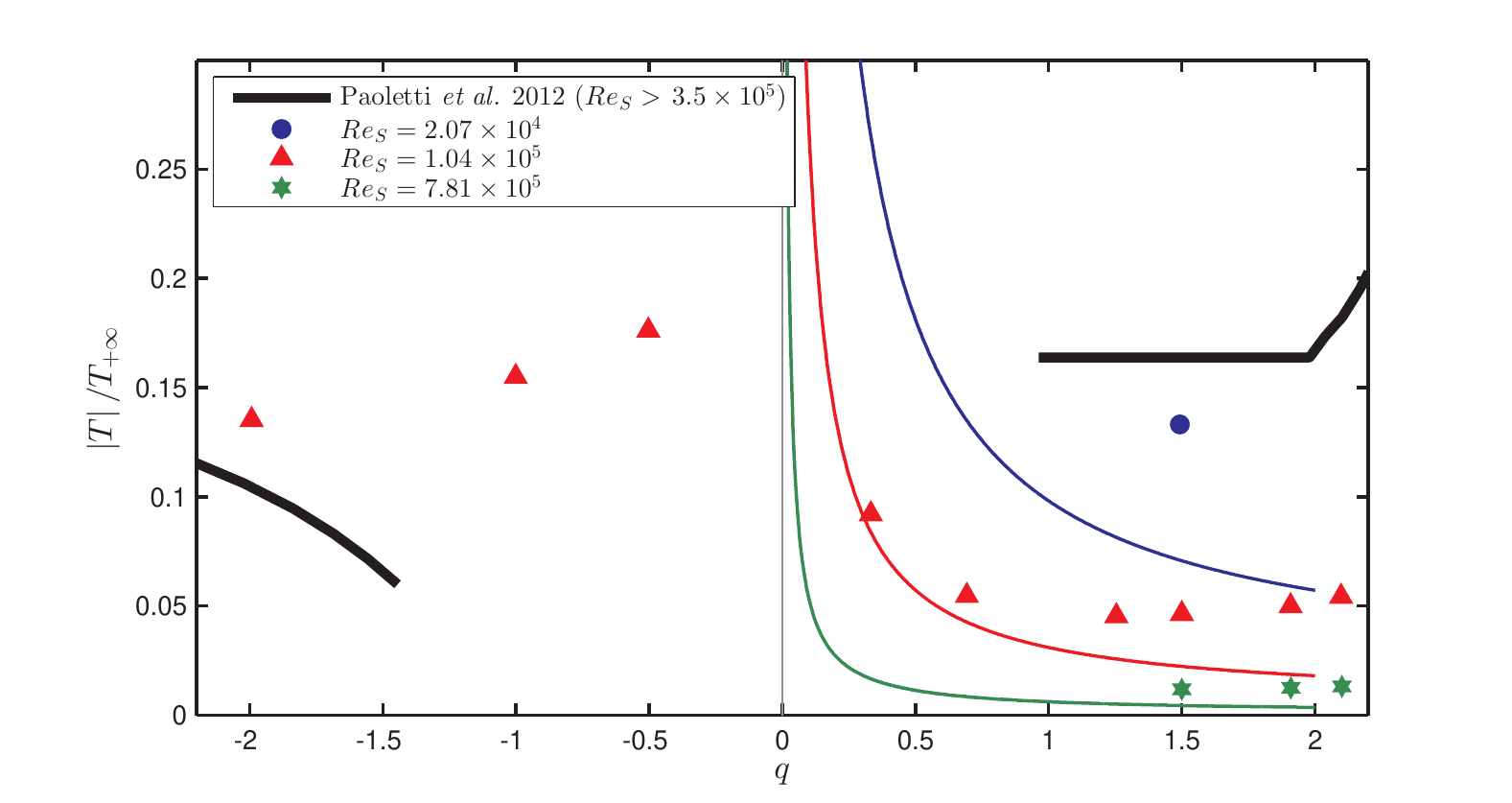}}
	\caption{Comparison of inner cylinder torque lower bounds calculated at $z / L = 0.209$ to the torques measured by \citet{paoletti_etal_AA_2012}.  Torques are normalized by the torque for pure inner rotation at the same $Re_S$. The scaling of \citet{paoletti_etal_AA_2012} for $Re_S > 3.5 \times 10^5$ is the thick black solid line. The torque ratio lower bounds obtained from the velocity profiles are the symbols, coded by $Re_S$.  The lower bound torques from the quasi-Keplerian flat inner angular momentum profile approximation in equation~(\ref{eqn:torque_lowerbound_quasiKeplerian}) for each $Re_S$ are the thin solid lines with the same colors as the symbols, which increase with decreasing $Re_S$.  Error bars are smaller than the symbols.}
	\label{fig:torque_lowerbounds}
\end{figure}

The torque lower bounds can be compared to the torque scaling that \citet{paoletti_etal_AA_2012} fit to the Maryland torque measurements in the Rayleigh-stable and unstable regimes \citep{paoletti_lathrop_PRL_2011} and the torque measurements on the apparatus presented in this paper in the unstable regime \citep{gils_etal_PRL_2011}.
The scaling was for the ratio of the torque on the inner cylinder to the torque $T_{+\infty}$ for pure inner rotation ($q = +\infty$) at the same $Re_S$, which in this paper was obtained from torque measurements in the very similar Maryland experiment \citep[equation~(9) in][]{lathrop_etal_PRA_1992}.
The lower bounds are compared to the torque scaling $Re_S > 3.5 \times 10^5$ \citep[equation~(12) in][]{paoletti_etal_AA_2012} in figure~\ref{fig:torque_lowerbounds}.

As the torque ratios must be positive, the torque scaling must start curving upwards on the sub-rotating regime side when approaching solid-body rotation at some $q > -1.5$ to avoid crossing zero.
The three sub-rotating regime torque lower bounds for $Re_S = 1.04 \times 10^5$ give $\left| T \right| / T_{+\infty}$ values that are larger than those for $Re_S > 3.5 \times 10^5$ \citep{paoletti_etal_AA_2012}.
Thus, the $\left| T \right| / T_{+\infty}$ scaling of \citet{paoletti_etal_AA_2012} must increase if extended to $Re_S = 1.04 \times 10^5$.

On the quasi-Keplerian side, comparisons can be made between our measurements at $Re_S = 7.81 \times 10^5$ to those of \citet{paoletti_etal_AA_2012} for $Re_S > 3.5 \times 10^5$.
Our $\left| T \right| / T_{+\infty}$ lower bounds from both the measured velocity profiles and the flat inner region approximation from equation~(\ref{eqn:torque_lowerbound_quasiKeplerian}), are considerably smaller than those of \citet{paoletti_etal_AA_2012}.
As our lower bounds only considered shear stress (diffusion), the difference in torques on the inner cylinder must be due to Reynolds stresses (advection) in the region of $\widetilde{r} \le 0.033$.
The divergence of the torque lower bound for the flat inner region angular momentum approximation as $q \to 0$ suggests that the flat quasi-Keplerian $\left| T \right| / T_{+\infty}$ scaling of \citet{paoletti_etal_AA_2012} will deviate from being flat if extended to $q < 1$, unless the inner region disappears or is distorted close to solid-body rotation.

\section{Summary and conclusions} \label{sec:conclusions}

In summary, azimuthal velocity profiles were obtained for several Rayleigh-stable (and one unstable) cylinder rotation rate ratios for the ranges $2.098 \ge q \ge 0.333$ and $-1.994 \le q \le -0.503$.
They were all done for $Re_S = 1.04 \times 10^5$, a few configurations at $Re_S = 7.81 \times 10^5$, and just the Keplerian configuration also at $Re_S = 2.07 \times 10^4$.
For all values of $q$, the profiles deviate from the laminar Taylor-Couette profile.
The deviation increases as solid-body rotation is approached ($q \rightarrow 0$) at fixed non-zero $Re_S$.
The deviation consists of a steepening of the normalized angular velocity $\widetilde{\omega}$ profile close to the inner cylinder for all $q$, and the flow in the outer parts of the gap approaching solid-body rotation with the outer cylinder and attached axial boundaries for the quasi-Keplerian regime.

For the sub-rotating regime, the flow exhibits super-rotation compared to both cylinders ($\omega > \Omega_o > \Omega_i$), except close to the inner cylinder.
As solid-body rotation is approached at fixed $Re_S = 1.04 \times 10^5$, the strength of the super-rotation increases, reaching $6\%$ of $\Omega_o - \Omega_i$ for $q = -0.503$, and the radial positions of the maximum of super-rotation and where the flow switched from $\Omega_i < \omega < \Omega_o$ to super-rotation moves closer to the inner cylinder.
The flow super-rotation must be sustained by inward angular momentum transport from the outer cylinder or axial boundaries.
To the best of our knowledge, flow super-rotation for $q < 0$ has not been previously observed in the literature.
This includes pure outer-rotation ($q = -\infty$) in our apparatus \citep{gils_etal_RevSciInst_2011} and in those of \citet{taylor_ProcRoySoc_1936_b}, \citet{wendt_1933}, and \citet{burin_czarnocki_JFM_2012}.

For the quasi-Keplerian regime, the specific angular momentum profiles show that the flow can be split into three regions across the gap: an inner region where the angular momentum profile is approximately flat, an outer region where the flow is close to solid-body rotation at $\Omega_o$, and a middle transition region between the two.
Starting near the Rayleigh line, the middle and outer regions are almost non-existent; and then as solid-body rotation is approached at fixed non-zero $Re_S$, the inner region shrinks while the outer region grows till the inner region is almost non-existent at $q = 0.333$.
As $Re_S$ is increased, the middle region shrinks.
We speculate that as $Re_S \to \infty$, the middle region will disappear and the profile will converge towards equation~(\ref{eqn:asymptotic_profile}) \citep[independently derived from dye injection observations by][]{dunst_JFM_1972}.
This model profile is a good approximation by $Re_S = 7.81 \times 10^5$.
The outer region, if present, likely transports little angular momentum, meaning that almost all of the angular momentum is transported to the axial boundaries.
Work is still needed to check how the mixing is achieved in the inner region and whether the inner region is turbulent.
One must also determine the exact nature of the flow in the middle and outer regions, and see whether the region pattern is found for other $\eta$ and $\Gamma$.
Measuring the separate torques on the inner cylinder, outer cylinder, and axial boundaries or doing very high resolution Particle Image Velocimetry as \citet{huisman_etal_PRL_2012} did for pure inner rotation on the same apparatus would be good ways to determine what fraction of the angular momentum goes to the axial boundaries versus the outer cylinder and elucidate the axial transport mechanism.
We do not see flow sub-rotation in our experiment except possibly for $q = 0.693$ and $0.333$, but $\widetilde{\omega} \approx 0$ within the measurement precision at those $q$.
However, \citet{kageyama_etal_JournPhysSocJapan_2004} and possibly \citet{edlund_ji_PRE_2014} found sub-rotating flow for $q > 0$.

The slope of the angular velocity profile at the inner cylinder is steeper than in laminar Taylor-Couette flow for both the sub-rotating and the quasi-Keplerian regimes.
Therefore, the torque required to rotate the inner cylinder must be larger than the laminar Taylor-Couette value, which supports the super-laminar torque measurements on the geometrically similar Maryland experiment \citep{paoletti_lathrop_PRL_2011}.
Due to not resolving the viscous sublayer on the inner cylinder, only lower bounds for the torque on the inner cylinder could be obtained via a viscous stress calculation.
The $\left| T \right| / T_{+\infty}$ lower bounds from the velocity measurements and the approximate asymptotic profile were compared to the $\left| T \right| / T_{+\infty}$ scaling found by \citet{paoletti_etal_AA_2012} for $Re_S > 3.5 \times 10^5$.
In the sub-rotating regime, their $\left| T \right| / T_{+\infty}$ scaling needs to be increased if extended to $Re_S = 10^5$ or towards solid-body rotation.
In the quasi-Keplerian regime, the comparison also shows that the bulk of the transport of angular momentum off the inner cylinder is by Reynolds stresses (advection) and the $\left| T \right| / T_{+\infty}$ scaling of \citet{paoletti_etal_AA_2012} may require modification as $q \rightarrow 0$.

Our velocity profiles provide experimental confirmation of the expectation that the Ekman pumping from the axial boundaries was what destabilized the flow in the Maryland experiment, which has a nearly identical geometry as our apparatus, in the Rayleigh-stable regime causing large super-laminar torques on the inner cylinder \citep{balbus_nature_2011,avila_PRL_2012,ji_balbus_PhysToday_2013,edlund_ji_PRE_2014}.
This work, combined with the work of \citet{avila_PRL_2012}, \citet{schartman_etal_AA_2012}, and \citet{edlund_ji_PRE_2014}, resolves the apparent discrepancy between the approximately laminar Taylor-Couette angular momentum transport in the wide-gap, low aspect-ratio experiments with axial boundaries split into rings rotating at speeds intermediate that of the cylinders such as the Princeton MRI and HTX experiments \citep{ji_etal_nature_2006,schartman_etal_AA_2012,edlund_ji_PRE_2014} and large super-laminar angular momentum transport in the medium-gap higher aspect-Maryland experiment with axial boundaries attached to the outer cylinder \citep{paoletti_lathrop_PRL_2011}.
Moreover, we found that the Ekman pumping from the axial boundaries does more than just destabilize the flow in the Rayleigh-stable regime when the axial boundaries are attached to the outer cylinder.
In the quasi-Keplerian regime, it causes the flow to be split radially into three regions and nearly all of the angular momentum to be transported to the axial boundaries instead of the outer cylinder when an outer region is present.
The Ekman pumping essentially causes the axial boundaries to become the primary sink of angular momentum.
In the sub-rotating regime, we discovered flow super-rotation, which is also likely due to the Ekman pumping.

Astrophysical accretion disks have open axial boundaries, which do not cause Ekman pumping, and are thought or assumed to have primarily radial transport of angular momentum \citep{zeldovich_ProcRoySoc_1981,richard_zahn_AA_1999,richard_thesis_2001,dubrulle_etal_POF_2005,ji_balbus_PhysToday_2013,monico_etal_JFM_2014b}.
Due to the strong Ekman pumping effects, including the primarily axial transport of angular momentum, Taylor-Couette flow with an aspect ratio up to $\Gamma \sim 10$ with no-slip axial boundaries attached to the outer cylinder is an imperfect model of accretion disks, especially with regard to stability.
Ideally, one would like to have axial boundaries that are free-slip or rotate at different rates along their radius such that they match the mean rotation rate of what the flow would be in the absence of axial boundaries, which may not be the laminar Taylor-Couette profile.

There are practical options available to experimental Taylor-Couette flow to mitigate the Ekman pumping and make a better model of accretion disks.
One practical way is to make an experiment where the aspect ratio is great enough that the axial tranport mechanism of the angular momentum saturates and Ekman pumping can no longer directly affect the flow near midheight.
However, tall experiments are difficult to handle and expensive to make, and work would be needed to ascertain whether indirect effects would still be a problem.
Another way, which has been followed by the Princeton group \citep{ji_etal_nature_2006,schartman_etal_AA_2012,edlund_ji_PRE_2014}, is to split the axial boundaries into rings that are rotated at speeds intermediate to those of the cylinders.
This reduces the strength of the Ekman pumping as well as better confining it to the axial boundaries.
Implementing the independently rotating rings is difficult and there is still Ekman pumping due to having only a finite number of independently rotating rings.
If the working fluid is a liquid, the top boundary can be made into an open boundary by having gas above it, reducing the Ekman pumping at the top by three orders of magnitude in the case of water and air, though it does introduce the problem of gravity waves on the top surface.
For the velocities that are used in the present experiments, air could be entrained by these waves.
Similarly, density-mismatched fluids such as mercury and water or stratification (e.g. salt solutions) can be used on the bottom boundary to confine the Ekman circulation near the bottom by reducing axial circulation, although this also introduces the problem of gravity waves and mixing which would destroy the stratification.
In order to accurately represent an accretion disk, one probably has to combine more than one of these methods.

\section*{Acknowledgements}

We would like to acknowledge helpful discussions and advice from Dennis P.M. van Gils, Siegfried Grossmann, Rodolfo Ostilla-M{\'o}nico, Daniel S. Zimmerman, Eric M. Edlund, and Hantao Ji.
We thank Hantao Ji for providing us the velocimetry data from \citet{kageyama_etal_JournPhysSocJapan_2004}.
We also acknowledge work and advice from the technicians Gert{-}Wim Bruggert, Martin Bos, and Bas Benschop; and financial support from the Technology Foundation STW of The Netherlands from an ERC Advanced Grant and the National Science Foundation of the USA (Grant No. NSF-DMR~0906109).

\bibliographystyle{hyperref_jfm}
\bibliography{TC}

\begin{thebibliography}{40}
\expandafter\ifx\csname natexlab\endcsname\relax\def\natexlab#1{#1}\fi

\bibitem[{Avila} {\em et~al.\/}(2011){Avila}, {Moxey}, {de Lozar}, {Avila},
  {Barkley} \& {Hof}]{avila_etal_science_2011}
{\sc {Avila}, K., {Moxey}, D., {de Lozar}, A., {Avila}, M., {Barkley}, D. \&
  {Hof}, B.} 2011 The Onset of Turbulence in Pipe Flow. {\em Science\/}
  \href{http://dx.doi.org/10.1126/science.1203223}{{\bf 333}, 192--6}.

\bibitem[{Avila}(2012)]{avila_PRL_2012}
{\sc {Avila}, M.} 2012 Stability and Angular-Momentum Transport of Fluid Flows
  between Corotating Cylinders. {\em Physical Review Letters\/}
  \href{http://dx.doi.org/10.1103/PhysRevLett.108.124501}{{\bf 108}~(12),
  124501}.

\bibitem[{Balbus}(2011)]{balbus_nature_2011}
{\sc {Balbus}, S.~A.} 2011 Fluid dynamics: A turbulent matter. {\em Nature\/}
  \href{http://dx.doi.org/10.1038/470475a}{{\bf 470}, 475--476}.

\bibitem[{Borrero-Echeverry} {\em et~al.\/}(2010){Borrero-Echeverry}, {Schatz}
  \& {Tagg}]{borrero_etal_PRE_2010}
{\sc {Borrero-Echeverry}, D., {Schatz}, M.~F. \& {Tagg}, R.} 2010 Transient
  turbulence in Taylor-Couette flow. {\em Physical Review E\/}
  \href{http://dx.doi.org/10.1103/PhysRevE.81.025301}{{\bf 81}~(2), 025301}.

\bibitem[{Burin} \& {Czarnocki}(2012)]{burin_czarnocki_JFM_2012}
{\sc {Burin}, M.~J. \& {Czarnocki}, C.~J.} 2012 Subcritical transition and
  spiral turbulence in circular Couette flow. {\em Journal of Fluid
  Mechanics\/} \href{http://dx.doi.org/10.1017/jfm.2012.323}{{\bf 709},
  106--122}.

\bibitem[{Coles}(1965)]{coles_JFM_1965}
{\sc {Coles}, D.} 1965 Transition in circular Couette flow. {\em Journal of
  Fluid Mechanics\/} \href{http://dx.doi.org/10.1017/S0022112065000241}{{\bf
  21}, 385--425}.

\bibitem[{Dubrulle} {\em et~al.\/}(2005{\natexlab{{\em a\/}}}){Dubrulle},
  {Dauchot}, {Daviaud}, {Longaretti}, {Richard} \&
  {Zahn}]{dubrulle_etal_POF_2005}
{\sc {Dubrulle}, B., {Dauchot}, O., {Daviaud}, F., {Longaretti}, P.-Y.,
  {Richard}, D. \& {Zahn}, J.-P.} 2005{\natexlab{{\em a\/}}} Stability and
  turbulent transport in Taylor-Couette flow from analysis of experimental
  data. {\em Physics of Fluids\/}
  \href{http://dx.doi.org/10.1063/1.2008999}{{\bf 17}~(9), 095103}.

\bibitem[{Dubrulle} {\em et~al.\/}(2005{\natexlab{{\em b\/}}}){Dubrulle},
  {Mari{\'e}}, {Normand}, {Richard}, {Hersant} \&
  {Zahn}]{dubrulle_etal_AA_2005}
{\sc {Dubrulle}, B., {Mari{\'e}}, L., {Normand}, C., {Richard}, D., {Hersant},
  F. \& {Zahn}, J.-P.} 2005{\natexlab{{\em b\/}}} {An hydrodynamic shear
  instability in stratified disks}. {\em A\&A\/}
  \href{http://dx.doi.org/10.1051/0004-6361:200400065}{{\bf 429}, 1--13}.

\bibitem[{Dunst}(1972)]{dunst_JFM_1972}
{\sc {Dunst}, M.} 1972 An experimental and analytical investigation of angular
  momentum exchange in a rotating fluid. {\em Journal of Fluid Mechanics\/}
  \href{http://dx.doi.org/10.1017/S0022112072001879}{{\bf 55}, 301--310}.

\bibitem[{Eckhardt} {\em et~al.\/}(2007){Eckhardt}, {Grossmann} \&
  {Lohse}]{eckhardt_etal_JFM_2007}
{\sc {Eckhardt}, B., {Grossmann}, S. \& {Lohse}, D.} 2007 Torque scaling in
  turbulent Taylor Couette flow between independently rotating cylinders. {\em
  Journal of Fluid Mechanics\/}
  \href{http://dx.doi.org/10.1017/S0022112007005629}{{\bf 581}, 221}.

\bibitem[{Edlund} \& {Ji}(2014)]{edlund_ji_PRE_2014}
{\sc {Edlund}, E.~M. \& {Ji}, H.} 2014 {Nonlinear stability of laboratory
  quasi-Keplerian flows}. {\em Physics Review E\/}
  \href{http://dx.doi.org/10.1103/PhysRevE.89.021004}{{\bf 89}~(2), 021004}.

\bibitem[{Grossmann}(2000)]{grossmann_rmp_2000}
{\sc {Grossmann}, S.} 2000 {The onset of shear flow turbulence}. {\em Reviews
  of Modern Physics\/} \href{http://dx.doi.org/10.1103/RevModPhys.72.603}{{\bf
  72}, 603--618}.

\bibitem[{Huisman} {\em et~al.\/}(2013){Huisman}, {Scharnowski}, {Cierpka},
  {K{\"a}hler}, {Lohse} \& {Sun}]{huisman_etal_PRL_2013}
{\sc {Huisman}, S.~G., {Scharnowski}, S., {Cierpka}, C., {K{\"a}hler}, C.~J.,
  {Lohse}, D. \& {Sun}, C.} 2013 {Logarithmic Boundary Layers in Strong
  Taylor-Couette Turbulence}. {\em Physical Review Letters\/}
  \href{http://dx.doi.org/10.1103/PhysRevLett.110.264501}{{\bf 110}~(26),
  264501}.

\bibitem[{Huisman} {\em et~al.\/}(2012{\natexlab{{\em a\/}}}){Huisman}, {van
  Gils}, {Grossmann}, {Sun} \& {Lohse}]{huisman_etal_PRL_2012}
{\sc {Huisman}, S.~G., {van Gils}, D.~P.~M., {Grossmann}, S., {Sun}, C. \&
  {Lohse}, D.} 2012{\natexlab{{\em a\/}}} Ultimate Turbulent Taylor-Couette
  Flow. {\em Physical Review Letters\/}
  \href{http://dx.doi.org/10.1103/PhysRevLett.108.024501}{{\bf 108}~(2),
  024501}.

\bibitem[{Huisman} {\em et~al.\/}(2012{\natexlab{{\em b\/}}}){Huisman}, {van
  Gils} \& {Sun}]{huisman_etal_EJMBF_2012}
{\sc {Huisman}, S.~G., {van Gils}, D.~P.~M. \& {Sun}, C.} 2012{\natexlab{{\em
  b\/}}} Applying laser Doppler anemometry inside a Taylor-Couette geometry
  using a ray-tracer to correct for curvature effects. {\em European Journal of
  Mechanics B Fluids\/}
  \href{http://dx.doi.org/10.1016/j.euromechflu.2012.03.013}{{\bf 36},
  115--119}.

\bibitem[Ji \& Balbus(2013)]{ji_balbus_PhysToday_2013}
{\sc Ji, Hantao \& Balbus, Steven} 2013 {Angular momentum transport in
  astrophysics and in the lab}. {\em Physics Today\/}
  \href{http://dx.doi.org/10.1063/PT.3.2081}{{\bf 66}~(8), 27}.

\bibitem[{Ji} {\em et~al.\/}(2006){Ji}, {Burin}, {Schartman} \&
  {Goodman}]{ji_etal_nature_2006}
{\sc {Ji}, H., {Burin}, M., {Schartman}, E. \& {Goodman}, J.} 2006 Hydrodynamic
  turbulence cannot transport angular momentum effectively in astrophysical
  disks. {\em Nature\/} \href{http://dx.doi.org/10.1038/nature05323}{{\bf 444},
  343--346}.

\bibitem[{Kageyama} {\em et~al.\/}(2004){Kageyama}, {Ji}, {Goodman}, {Chen} \&
  {Shoshan}]{kageyama_etal_JournPhysSocJapan_2004}
{\sc {Kageyama}, A., {Ji}, H., {Goodman}, J., {Chen}, F. \& {Shoshan}, E.} 2004
  Numerical and Experimental Investigation of Circulation in Short Cylinders.
  {\em Journal of the Physical Society of Japan\/}
  \href{http://dx.doi.org/10.1143/JPSJ.73.2424}{{\bf 73}, 2424}.

\bibitem[Landau \& Lifshitz(1987)]{landau_lifshitz_fluids_1987}
{\sc Landau, L.D. \& Lifshitz, E.M.} 1987 {\em Fluid Mechanics\/}, 2nd edn.,
  {\em Course of Theoretical Physics\/}, vol.~6. Pergamon Press.

\bibitem[{Lathrop} {\em et~al.\/}(1992){Lathrop}, {Fineberg} \&
  {Swinney}]{lathrop_etal_PRA_1992}
{\sc {Lathrop}, D.~P., {Fineberg}, J. \& {Swinney}, H.~L.} 1992 Transition to
  shear-driven turbulence in Couette-Taylor flow. {\em Physical Review A\/}
  \href{http://dx.doi.org/10.1103/PhysRevA.46.6390}{{\bf 46}, 6390--6405}.

\bibitem[{Le Bars} \& {Le Gal}(2007)]{bars_gal_PRE_2007}
{\sc {Le Bars}, M. \& {Le Gal}, P.} 2007 {Experimental Analysis of the
  Stratorotational Instability in a Cylindrical Couette Flow}. {\em Physical
  Review Letters\/} \href{http://dx.doi.org/10.1103/PhysRevLett.99.064502}{{\bf
  99}~(6), 064502}.

\bibitem[{Le Diz{\`e}s} \& {Riedinger}(2010)]{ledizes_riedinger_JFM_2010}
{\sc {Le Diz{\`e}s}, S. \& {Riedinger}, X.} 2010 {The strato-rotational
  instability of Taylor-Couette and Keplerian flows}. {\em Journal of Fluid
  Mechanics\/} \href{http://dx.doi.org/10.1017/S0022112010002624}{{\bf 660},
  147--161}.

\bibitem[Maretzke {\em et~al.\/}(2014)Maretzke, Hof \&
  Avila]{maretzke_hof_avila_JFM_2014}
{\sc Maretzke, Simon, Hof, Bj{\"o}rn \& Avila, Marc} 2014 {Transient growth in
  linearly stable Taylor-Couette flows}. {\em Journal of Fluid Mechanics\/}
  \href{http://dx.doi.org/10.1017/jfm.2014.12}{{\bf 742}, 254--290}.

\bibitem[{Ostilla-M{\'o}nico} {\em et~al.\/}(2014){Ostilla-M{\'o}nico},
  {Verzicco}, {Grossmann} \& {Lohse}]{monico_etal_JFM_2014b}
{\sc {Ostilla-M{\'o}nico}, R., {Verzicco}, R., {Grossmann}, S. \& {Lohse}, D.}
  2014 Turbulence decay towards the linearly stable regime of Taylor-Couette
  flow. {\em Journal of Fluid Mechanics\/}
  \href{http://dx.doi.org/10.1017/jfm.2014.242}{{\bf 748}, 3}.

\bibitem[{Paoletti} \& {Lathrop}(2011)]{paoletti_lathrop_PRL_2011}
{\sc {Paoletti}, M.~S. \& {Lathrop}, D.~P.} 2011 Angular Momentum Transport in
  Turbulent Flow between Independently Rotating Cylinders. {\em Physical Review
  Letters\/} \href{http://dx.doi.org/10.1103/PhysRevLett.106.024501}{{\bf
  106}~(2), 024501}.

\bibitem[{Paoletti} {\em et~al.\/}(2012){Paoletti}, {van Gils}, {Dubrulle},
  {Sun}, {Lohse} \& {Lathrop}]{paoletti_etal_AA_2012}
{\sc {Paoletti}, M.~S., {van Gils}, D.~P.~M., {Dubrulle}, B., {Sun}, C.,
  {Lohse}, D. \& {Lathrop}, D.~P.} 2012 Angular momentum transport and
  turbulence in laboratory models of Keplerian flows. {\em A\&A\/}
  \href{http://dx.doi.org/10.1051/0004-6361/201118511}{{\bf 547}, A64}.

\bibitem[Rayleigh(1917)]{rayleigh_prsa_1917}
{\sc Rayleigh, Lord} 1917 On the Dynamics of Revolving Fluids. {\em Proceedings
  of the Royal Society of London Series A\/}
  \href{http://dx.doi.org/10.1098/rspa.1917.0010}{{\bf 93}~(648), 148--154}.

\bibitem[Richard(2001)]{richard_thesis_2001}
{\sc Richard, Denis} 2001 Instabilit{\'e}s hydrodynamiques dans les
  {\'e}coulements en rotation diff{\'e}rentielle. Ph.d. thesis,
  \href{http://adsabs.harvard.edu/abs/2001PhDT.........9R}{Universit{\'e}
  Paris-Diderot - Paris VII}.

\bibitem[{Richard} \& {Zahn}(1999)]{richard_zahn_AA_1999}
{\sc {Richard}, D. \& {Zahn}, J.-P.} 1999 {Turbulence in differentially
  rotating flows. What can be learned from the Couette-Taylor experiment}. {\em
  A\&A\/} \href{http://adsabs.harvard.edu/abs/1999A%26A...347..734R}{{\bf 347},
  734--738}.

\bibitem[{Schartman} {\em et~al.\/}(2012){Schartman}, {Ji}, {Burin} \&
  {Goodman}]{schartman_etal_AA_2012}
{\sc {Schartman}, E., {Ji}, H., {Burin}, M.~J. \& {Goodman}, J.} 2012 Stability
  of quasi-Keplerian shear flow in a laboratory experiment. {\em A\&A\/}
  \href{http://dx.doi.org/10.1051/0004-6361/201016252}{{\bf 543}, A94}.

\bibitem[Schlichting(1979)]{schlichting_boundaryLayerTheory_1979}
{\sc Schlichting, Hermann~T.} 1979 {\em Boundary Layer Theory\/}. McGraw-Hill
  Science/Engineering/Math.

\bibitem[{Shi} {\em et~al.\/}(2013){Shi}, {Avila} \& {Hof}]{shi_etal_PRL_2013}
{\sc {Shi}, L., {Avila}, M. \& {Hof}, B.} 2013 Scale Invariance at the Onset of
  Turbulence in Couette Flow. {\em Physical Review Letters\/}
  \href{http://dx.doi.org/10.1103/PhysRevLett.110.204502}{{\bf 110}~(20),
  204502}.

\bibitem[{Taylor}(1923)]{taylor_prsa_1923}
{\sc {Taylor}, G.~I.} 1923 {Stability of a Viscous Liquid Contained between Two
  Rotating Cylinders}. {\em Royal Society of London Philosophical Transactions
  Series A\/} \href{http://dx.doi.org/10.1098/rsta.1923.0008}{{\bf 223},
  289--343}.

\bibitem[Taylor(1936{\natexlab{{\em a\/}}})]{taylor_ProcRoySoc_1936_a}
{\sc Taylor, G.~I.} 1936{\natexlab{{\em a\/}}} Fluid Friction between Rotating
  Cylinders. I. Torque Measurements. {\em Proceedings of the Royal Society of
  London Series A\/} \href{http://dx.doi.org/10.1098/rspa.1936.0215}{{\bf
  157}~(892), 546--564}.

\bibitem[Taylor(1936{\natexlab{{\em b\/}}})]{taylor_ProcRoySoc_1936_b}
{\sc Taylor, G.~I.} 1936{\natexlab{{\em b\/}}} {Fluid Friction between Rotating
  Cylinders. II. Distribution of Velocity between Concentric Cylinders when
  Outer One Is Rotating and Inner One Is at Rest}. {\em Proceedings of the
  Royal Society of London Series A\/}
  \href{http://dx.doi.org/10.1098/rspa.1936.0216}{{\bf 157}, 565--578}.

\bibitem[{van Gils} {\em et~al.\/}(2011{\natexlab{{\em a\/}}}){van Gils},
  {Bruggert}, {Lathrop}, {Sun} \& {Lohse}]{gils_etal_RevSciInst_2011}
{\sc {van Gils}, D.~P.~M., {Bruggert}, G.-W., {Lathrop}, D.~P., {Sun}, C. \&
  {Lohse}, D.} 2011{\natexlab{{\em a\/}}} The Twente turbulent Taylor-Couette
  (T3C) facility: Strongly turbulent (multiphase) flow between two
  independently rotating cylinders. {\em Review of Scientific Instruments\/}
  \href{http://dx.doi.org/10.1063/1.3548924}{{\bf 82}~(2), 025105}.

\bibitem[{van Gils} {\em et~al.\/}(2011{\natexlab{{\em b\/}}}){van Gils},
  {Huisman}, {Bruggert}, {Sun} \& {Lohse}]{gils_etal_PRL_2011}
{\sc {van Gils}, D.~P.~M., {Huisman}, S.~G., {Bruggert}, G.-W., {Sun}, C. \&
  {Lohse}, D.} 2011{\natexlab{{\em b\/}}} Torque Scaling in Turbulent
  Taylor-Couette Flow with Co- and Counterrotating Cylinders. {\em Physical
  Review Letters\/}
  \href{http://dx.doi.org/10.1103/PhysRevLett.106.024502}{{\bf 106}~(2),
  024502}.

\bibitem[{van Gils} {\em et~al.\/}(2012){van Gils}, {Huisman}, {Grossmann},
  {Sun} \& {Lohse}]{gils_etal_JFM_2012}
{\sc {van Gils}, D.~P.~M., {Huisman}, S.~G., {Grossmann}, S., {Sun}, C. \&
  {Lohse}, D.} 2012 Optimal Taylor-Couette turbulence. {\em Journal of Fluid
  Mechanics\/} \href{http://dx.doi.org/10.1017/jfm.2012.236}{{\bf 706},
  118--149}.

\bibitem[Wendt(1933)]{wendt_1933}
{\sc Wendt, Fritz} 1933 Turbulente Str{\"o}mungen zwischen zwei rotierenden
  konaxialen Zylindern. {\em Ingenieurs et architectes suisses\/}
  \href{http://dx.doi.org/10.1007/BF02084936}{{\bf 4}, 577--595}.

\bibitem[{Zeldovich}(1981)]{zeldovich_ProcRoySoc_1981}
{\sc {Zeldovich}, Y.~B.} 1981 {On the Friction of Fluids Between Rotating
  Cylinders}. {\em Proceedings of the Royal Society of London Series A\/}
  \href{http://dx.doi.org/10.1098/rspa.1981.0024}{{\bf 374}, 299--312}.

\end{thebibliography}

\end{document}